\PassOptionsToPackage{table}{xcolor} % for table row colors
\documentclass[acmsmall]{acmart}

\AtBeginDocument{%
  \providecommand\BibTeX{{%
    \normalfont B\kern-0.5em{\scshape i\kern-0.25em b}\kern-0.8em\TeX}}}

\setcopyright{acmlicensed}
\copyrightyear{2026}
\acmYear{2026}
\acmDOI{XXXXXXX.XXXXXXX}

\acmJournal{CSUR}
\acmVolume{XX}
\acmNumber{XX}
\acmArticle{XXX}
\acmMonth{6}
\usepackage{tikz}
\usetikzlibrary{trees}
\usepackage{forest}
\usepackage{adjustbox}
\usepackage{lscape}
\usepackage{tikz}
\usepackage{multirow}
\usepackage{lscape}
\usepackage{longtable}
\usepackage{pifont} % for checkmark symbol
\usepackage{xcolor}  % for revision highlighting
\definecolor{rowgray}{gray}{0.92} % for table row colors
\newcommand{\rev}[1]{\textcolor{black}{#1}}  % command for revision text
\newcommand{\minrev}[1]{\textcolor{black}{#1}}

\begin{document}

\title{Dynamic Function Configuration and its Management in Serverless Computing: A Taxonomy and Future Directions}

\author{Siddharth Agarwal}
\orcid{0000-0003-0944-8314}
\email{siddhartha@student.unimelb.edu.au}
\affiliation{%
  \institution{Quantum Cloud Computing and Distributed Systems (qCLOUDS) Lab, School of Computing and Information Systems, The University of Melbourne}
  \streetaddress{700 Swanston Street}
  \city{Parkville}
  \state{Victoria}
  \country{Australia}
  \postcode{3010}
}

\author{Maria A. Rodriguez}
\orcid{0000-0002-2831-8526}
\email{maria.read@unimelb.edu.au}
\affiliation{%
  \institution{Quantum Cloud Computing and Distributed Systems (qCLOUDS) Lab, School of Computing and Information Systems, The University of Melbourne}
  \streetaddress{700 Swanston Street}
  \city{Parkville}
  \state{Victoria}
  \country{Australia}
  \postcode{3010}
}

\author{Rajkumar Buyya}
\orcid{0000-0001-9754-6496}
\email{rbuyya@unimelb.edu.au}
\affiliation{%
  \institution{Quantum Cloud Computing and Distributed Systems (qCLOUDS) Lab, School of Computing and Information Systems, The University of Melbourne}
  \streetaddress{700 Swanston Street}
  \city{Parkville}
  \state{Victoria}
  \country{Australia}
  \postcode{3010}
}

\renewcommand{\shortauthors}{Agarwal, Rodriguez and Buyya}

\begin{abstract}
The serverless cloud computing model offers a framework where the service provider abstracts the underlying infrastructure management from developers. In this serverless model, Function-as-a-Service (FaaS) provides an event-driven, function-oriented computing service characterised by fine-grained, usage-based pricing that eliminates cost for idle resources. Platforms like AWS Lambda, Azure Functions, and Cloud Run Functions require developers to configure their \textit{function(s)} with minimum operational resources for its successful execution. This resource allocation influences both the operational expense and the performance quality of these functions. However, a noticeable lack of platform transparency forces developers to rely on expert knowledge or experience-based ad-hoc decisions to request desired function resources. This makes optimal resource configuration a non-trivial task while adhering to performance constraints. Furthermore, while commercial platforms often scale resources like CPU and network bandwidth proportional to memory, open-source frameworks permit independent configuration of function resources, introducing additional complexity for developers aiming to optimise their functions. These complexities have directed researchers to resolve developer challenges and advance towards an efficient \textit{server-less} execution model. In this article, we identify different aspects of resource configuration techniques in FaaS settings and propose a taxonomy of factors that influence function design, configuration, run-time cost, and performance guarantees from both a developer's and service provider's perspective. We conduct an analysis of existing literature on resource configuration in FaaS and present a comprehensive review of current studies on function configuration. This article also identifies existing research gaps and suggests future research directions for enhancing function configuration and strengthening the capabilities of serverless computing environments to drive its broader adoption.

\end{abstract}




\maketitle

\section{Introduction}

% \\Brief introduction
In the modern cloud computing era, serverless computing has seen a rapid adoption in the industry owing to its reduced resource management burden for developers. Serverless computing emerges as the latest cloud-native offering that encourages the idea of loosely integrated provider-managed services with a usage-based resource pricing model. Over the years, \textit{serverless} has evolved from the idea of running code without the need for server management to being promoted as a strategic mindset, a practice that allows organisations to focus on business value rather than worrying about underlying technology \cite{awsReInvent23}. 

% \\Brief introduction to Faas and Baas + When was the model launched
With the serverless model, service providers abstract away the underlying resources from developers and expose services to put together a cloud-native solution. The serverless paradigm encompasses two service models, namely Function-as-a-Service (FaaS) and Backend-as-a-Service (BaaS). FaaS offers an event-driven function-oriented compute service that hides the underlying servers. A \textit{function} is a code fragment that is purpose-built and associated with operational resources like compute power and memory for its execution. BaaS, on the other hand, constitutes the complementary managed services such as storage, database, and networking that complete the serverless application development suite. In 2014, Amazon Web Services (AWS) introduced AWS Lambda \cite{awsLambda} as their FaaS offering, and since then, many Cloud Service Providers (CSPs) have launched their own FaaS platforms, including Azure Functions \cite{azureFunctions}, Cloud Run Functions \rev{(formerly Google Cloud Functions \cite{gcf})}\cite{runFunctions} and Oracle Functions \cite{oracleFunctions}, along with many open-source frameworks such as OpenFaaS \cite{openfaas} and Apache OpenWhisk \cite{openwhisk}. The FaaS model maintains unique attributes such as event-driven ephemeral execution, statelessness, and no idle resource cost. Therefore, its application can be found in a variety of use cases such as event-driven websites \cite{ServerlessLand}, video streaming platforms \cite{ServerlessVideo}, multimedia processing \cite{NetflixAWS}, and CI/CD pipelines \cite{CapitalOne}.

% \\FaaS and resource configuration importance
FaaS promotes rapid application development by shifting underlying server management activities such as resource planning, provisioning, scheduling, and system maintenance to the service provider. However, it still necessitates developers to configure their \textit{functions} with appropriate settings, such as operational memory, ephemeral storage or function timeout. These different resource settings are known to significantly affect a function's performance and cost \cite{TimeCostMemoryConfig}\cite{COSE.V2}\cite{SLAM}\cite{StepConf}, where the set of configurable settings may vary across different FaaS providers. For example, in AWS Lambda, users define the amount of memory to be allocated to a function's execution, while the platform transparently allocates other resources in proportion to the memory configuration. On the other hand, OpenFaaS offers a fine-grained control over resource configurations by allowing users to define memory and CPU allocations independently. FaaS platforms often scale functions based on the static resource configuration provided during creation and expect a consistent performance across the workload. However, studies \cite{FireFace}\cite{ParrotFish}\cite{OFC} have identified that application and workload-related features like input characteristics, workload demand and seasonality, inter-workflow dependency, function capacity, and resource-affinity, also significantly impact a function's resource requirements. This leaves function configuration as a non-trivial and cumbersome task that requires meticulous function analysis and a trade-off of various parameters to balance run-time cost with performance. To this end, most developers often resort to default function settings \cite{Datadog} or employ experience-based ad-hoc configuration decisions \cite{CPUTAMS} while expecting a stable function performance.

\begin{landscape}
\scriptsize
\begin{longtable}{|c|c|ccc|ccccc|ccc|c|}
\caption{A categorisation of related surveys in Serverless Computing }
\label{tab:related-surveys}\\
\hline
\multicolumn{1}{|c|}{\multirow{2}{*}{Work}} & \multicolumn{1}{c|}{\multirow{2}{*}{Year}} & \multicolumn{3}{c|}{Resource Management} & \multicolumn{5}{c|}{Challenges \& Opportunities} & \multicolumn{3}{c|}{Workload} & \multicolumn{1}{c|}{\multirow{2}{*}{General Overview}} \\ \cline{3-13} 
\multicolumn{1}{|c|}{} & \multicolumn{1}{c|}{} & \multicolumn{1}{c|}{Scheduling} & \multicolumn{1}{c|}{Scaling} & \multicolumn{1}{c|}{Configuration} & \multicolumn{1}{c|}{Infrastructure} & \multicolumn{1}{c|}{Cold Start} & \multicolumn{1}{c|}{Cost} & \multicolumn{1}{c|}{Performance} & \multicolumn{1}{c|}{Security} & \multicolumn{1}{c|}{Characterisation} & \multicolumn{1}{c|}{Development} & \multicolumn{1}{c|}{Workflows} &  \\ \hline
\endhead
%
% ROW 1 — grey
\rowcolor{rowgray}
\cite{agbaje2022serverless} & 2022 & \multicolumn{1}{c|}{\ding{55}} & \multicolumn{1}{c|}{\ding{55}} & \ding{55} & \multicolumn{1}{c|}{\ding{55}} & \multicolumn{1}{c|}{\ding{55}} & \multicolumn{1}{c|}{\ding{51}} & \multicolumn{1}{c|}{\ding{55}} & \ding{55} & \multicolumn{1}{c|}{\ding{55}} & \multicolumn{1}{c|}{\ding{51}} & \ding{55} & \ding{55} \\ \hline
% ROW 2 — white
\cite{scheduling2018} & 2018 & \multicolumn{1}{c|}{\ding{51}} & \multicolumn{1}{c|}{\ding{55}} & \ding{55} & \multicolumn{1}{c|}{\ding{55}} & \multicolumn{1}{c|}{\ding{55}} & \multicolumn{1}{c|}{\ding{55}} & \multicolumn{1}{c|}{\ding{55}} & \ding{55} & \multicolumn{1}{c|}{\ding{55}} & \multicolumn{1}{c|}{\ding{55}} & \ding{55} & \ding{55} \\ \hline
% ROW 3 — grey
\rowcolor{rowgray}
\cite{BaldiniChalOpp2017} & 2017 & \multicolumn{1}{c|}{\ding{55}} & \multicolumn{1}{c|}{\ding{55}} & \ding{55} & \multicolumn{1}{c|}{\ding{55}} & \multicolumn{1}{c|}{\ding{55}} & \multicolumn{1}{c|}{\ding{55}} & \multicolumn{1}{c|}{\ding{55}} & \ding{55} & \multicolumn{1}{c|}{\ding{55}} & \multicolumn{1}{c|}{\ding{55}} & \ding{55} & \ding{51} \\ \hline
% ROW 4 — white
\cite{eismannApp2022} & 2022 & \multicolumn{1}{c|}{\ding{55}} & \multicolumn{1}{c|}{\ding{55}} & \ding{55} & \multicolumn{1}{c|}{\ding{55}} & \multicolumn{1}{c|}{\ding{55}} & \multicolumn{1}{c|}{\ding{55}} & \multicolumn{1}{c|}{\ding{55}} & \ding{55} & \multicolumn{1}{c|}{\ding{51}} & \multicolumn{1}{c|}{\ding{51}} & \ding{51} & \ding{55} \\ \hline
% ROW 5 — grey
\rowcolor{rowgray}
\cite{golec2024cold} & 2024 & \multicolumn{1}{c|}{\ding{55}} & \multicolumn{1}{c|}{\ding{55}} & \ding{55} & \multicolumn{1}{c|}{\ding{55}} & \multicolumn{1}{c|}{\ding{51}} & \multicolumn{1}{c|}{\ding{55}} & \multicolumn{1}{c|}{\ding{55}} & \ding{55} & \multicolumn{1}{c|}{\ding{55}} & \multicolumn{1}{c|}{\ding{55}} & \ding{55} & \ding{55} \\ \hline
% ROW 6 — white
\cite{resMan2023} & 2023 & \multicolumn{1}{c|}{\ding{51}} & \multicolumn{1}{c|}{\ding{55}} & \ding{55} & \multicolumn{1}{c|}{\ding{55}} & \multicolumn{1}{c|}{\ding{55}} & \multicolumn{1}{c|}{\ding{55}} & \multicolumn{1}{c|}{\ding{55}} & \ding{55} & \multicolumn{1}{c|}{\ding{55}} & \multicolumn{1}{c|}{\ding{55}} & \ding{55} & \ding{51} \\ \hline
% ROW 7 — grey
\rowcolor{rowgray}
\cite{economy2024} & 2024 & \multicolumn{1}{c|}{\ding{55}} & \multicolumn{1}{c|}{\ding{55}} & \ding{55} & \multicolumn{1}{c|}{\ding{55}} & \multicolumn{1}{c|}{\ding{55}} & \multicolumn{1}{c|}{\ding{51}} & \multicolumn{1}{c|}{\ding{55}} & \ding{55} & \multicolumn{1}{c|}{\ding{55}} & \multicolumn{1}{c|}{\ding{55}} & \ding{55} & \ding{55} \\ \hline
% ROW 8 — white
\cite{hassan2021survey} & 2021 & \multicolumn{1}{c|}{\ding{55}} & \multicolumn{1}{c|}{\ding{55}} & \ding{55} & \multicolumn{1}{c|}{\ding{55}} & \multicolumn{1}{c|}{\ding{55}} & \multicolumn{1}{c|}{\ding{55}} & \multicolumn{1}{c|}{\ding{55}} & \ding{55} & \multicolumn{1}{c|}{\ding{55}} & \multicolumn{1}{c|}{\ding{55}} & \ding{55} & \ding{51} \\ \hline
% ROW 9 — grey
\rowcolor{rowgray}
\cite{hellerstein2018serverless} & 2018 & \multicolumn{1}{c|}{\ding{55}} & \multicolumn{1}{c|}{\ding{55}} & \ding{55} & \multicolumn{1}{c|}{\ding{55}} & \multicolumn{1}{c|}{\ding{55}} & \multicolumn{1}{c|}{\ding{55}} & \multicolumn{1}{c|}{\ding{55}} & \ding{55} & \multicolumn{1}{c|}{\ding{55}} & \multicolumn{1}{c|}{\ding{55}} & \ding{55} & \ding{51} \\ \hline
% ROW 10 — white
\cite{berkleyView2019} & 2019 & \multicolumn{1}{c|}{\ding{55}} & \multicolumn{1}{c|}{\ding{55}} & \ding{55} & \multicolumn{1}{c|}{\ding{55}} & \multicolumn{1}{c|}{\ding{55}} & \multicolumn{1}{c|}{\ding{55}} & \multicolumn{1}{c|}{\ding{55}} & \ding{55} & \multicolumn{1}{c|}{\ding{55}} & \multicolumn{1}{c|}{\ding{55}} & \ding{55} & \ding{51} \\ \hline
% ROW 11 — grey
\rowcolor{rowgray}
\cite{whatSCis2023} & 2023 & \multicolumn{1}{c|}{\ding{55}} & \multicolumn{1}{c|}{\ding{55}} & \ding{55} & \multicolumn{1}{c|}{\ding{55}} & \multicolumn{1}{c|}{\ding{55}} & \multicolumn{1}{c|}{\ding{55}} & \multicolumn{1}{c|}{\ding{55}} & \ding{55} & \multicolumn{1}{c|}{\ding{55}} & \multicolumn{1}{c|}{\ding{55}} & \ding{55} & \ding{51} \\ \hline
% ROW 12 — white
\cite{LannurienChalOpp2023} & 2023 & \multicolumn{1}{c|}{\ding{55}} & \multicolumn{1}{c|}{\ding{55}} & \ding{55} & \multicolumn{1}{c|}{\ding{55}} & \multicolumn{1}{c|}{\ding{55}} & \multicolumn{1}{c|}{\ding{55}} & \multicolumn{1}{c|}{\ding{55}} & \ding{55} & \multicolumn{1}{c|}{\ding{55}} & \multicolumn{1}{c|}{\ding{55}} & \ding{55} & \ding{51} \\ \hline
% ROW 13 — grey
\rowcolor{rowgray}
\cite{evalProd2018} & 2018 & \multicolumn{1}{c|}{\ding{55}} & \multicolumn{1}{c|}{\ding{55}} & \ding{55} & \multicolumn{1}{c|}{\ding{55}} & \multicolumn{1}{c|}{\ding{55}} & \multicolumn{1}{c|}{\ding{55}} & \multicolumn{1}{c|}{\ding{55}} & \ding{55} & \multicolumn{1}{c|}{\ding{55}} & \multicolumn{1}{c|}{\ding{51}} & \ding{55} & \ding{55} \\ \hline
% ROW 14 — white
\cite{archTechnicalPrimer2022} & 2022 & \multicolumn{1}{c|}{\ding{55}} & \multicolumn{1}{c|}{\ding{55}} & \ding{55} & \multicolumn{1}{c|}{\ding{55}} & \multicolumn{1}{c|}{\ding{55}} & \multicolumn{1}{c|}{\ding{55}} & \multicolumn{1}{c|}{\ding{55}} & \ding{51} & \multicolumn{1}{c|}{\ding{55}} & \multicolumn{1}{c|}{\ding{55}} & \ding{55} & \ding{55} \\ \hline
% ROW 15 — grey
\rowcolor{rowgray}
\cite{economy2020} & 2020 & \multicolumn{1}{c|}{\ding{55}} & \multicolumn{1}{c|}{\ding{55}} & \ding{55} & \multicolumn{1}{c|}{\ding{55}} & \multicolumn{1}{c|}{\ding{55}} & \multicolumn{1}{c|}{\ding{51}} & \multicolumn{1}{c|}{\ding{55}} & \ding{55} & \multicolumn{1}{c|}{\ding{55}} & \multicolumn{1}{c|}{\ding{55}} & \ding{55} & \ding{55} \\ \hline
% ROW 16 — white
\cite{evalMicroSvc2018} & 2018 & \multicolumn{1}{c|}{\ding{55}} & \multicolumn{1}{c|}{\ding{55}} & \ding{55} & \multicolumn{1}{c|}{\ding{51}} & \multicolumn{1}{c|}{\ding{55}} & \multicolumn{1}{c|}{\ding{55}} & \multicolumn{1}{c|}{\ding{51}} & \ding{55} & \multicolumn{1}{c|}{\ding{55}} & \multicolumn{1}{c|}{\ding{55}} & \ding{55} & \ding{55} \\ \hline
% ROW 17 — grey
\rowcolor{rowgray}
\cite{schedAuto2023} & 2023 & \multicolumn{1}{c|}{\ding{51}} & \multicolumn{1}{c|}{\ding{51}} & \ding{55} & \multicolumn{1}{c|}{\ding{55}} & \multicolumn{1}{c|}{\ding{55}} & \multicolumn{1}{c|}{\ding{55}} & \multicolumn{1}{c|}{\ding{55}} & \ding{55} & \multicolumn{1}{c|}{\ding{55}} & \multicolumn{1}{c|}{\ding{55}} & \ding{55} & \ding{55} \\ \hline
% ROW 18 — white
\cite{anupamaSurvey} & 2022 & \multicolumn{1}{c|}{\ding{51}} & \multicolumn{1}{c|}{\ding{51}} & \ding{55} & \multicolumn{1}{c|}{\ding{55}} & \multicolumn{1}{c|}{\ding{55}} & \multicolumn{1}{c|}{\ding{55}} & \multicolumn{1}{c|}{\ding{55}} & \ding{55} & \multicolumn{1}{c|}{\ding{51}} & \multicolumn{1}{c|}{\ding{55}} & \ding{55} & \ding{55} \\ \hline
% ROW 19 — grey
\rowcolor{rowgray}
\cite{latency2022} & 2022 & \multicolumn{1}{c|}{\ding{55}} & \multicolumn{1}{c|}{\ding{55}} & \ding{55} & \multicolumn{1}{c|}{\ding{51}} & \multicolumn{1}{c|}{\ding{55}} & \multicolumn{1}{c|}{\ding{55}} & \multicolumn{1}{c|}{\ding{55}} & \ding{55} & \multicolumn{1}{c|}{\ding{55}} & \multicolumn{1}{c|}{\ding{55}} & \ding{55} & \ding{55} \\ \hline
% ROW 20 — white
\cite{whatSCis2021} & 2021 & \multicolumn{1}{c|}{\ding{55}} & \multicolumn{1}{c|}{\ding{55}} & \ding{55} & \multicolumn{1}{c|}{\ding{55}} & \multicolumn{1}{c|}{\ding{55}} & \multicolumn{1}{c|}{\ding{55}} & \multicolumn{1}{c|}{\ding{55}} & \ding{55} & \multicolumn{1}{c|}{\ding{55}} & \multicolumn{1}{c|}{\ding{55}} & \ding{55} & \ding{51} \\ \hline
% ROW 21 — grey
\rowcolor{rowgray}
\cite{shafieiChalOpp2022} & 2022 & \multicolumn{1}{c|}{\ding{55}} & \multicolumn{1}{c|}{\ding{55}} & \ding{55} & \multicolumn{1}{c|}{\ding{55}} & \multicolumn{1}{c|}{\ding{55}} & \multicolumn{1}{c|}{\ding{55}} & \multicolumn{1}{c|}{\ding{55}} & \ding{55} & \multicolumn{1}{c|}{\ding{55}} & \multicolumn{1}{c|}{\ding{55}} & \ding{55} & \ding{51} \\ \hline
% ROW 22 — white
\cite{shahradArchitecture2019} & 2019 & \multicolumn{1}{c|}{\ding{55}} & \multicolumn{1}{c|}{\ding{55}} & & \multicolumn{1}{c|}{\ding{51}} & \multicolumn{1}{c|}{\ding{55}} & \multicolumn{1}{c|}{\ding{55}} & \multicolumn{1}{c|}{\ding{55}} & \ding{55} & \multicolumn{1}{c|}{\ding{55}} & \multicolumn{1}{c|}{\ding{55}} & \ding{55} & \ding{55} \\ \hline
% ROW 23 — grey
\rowcolor{rowgray}
\cite{pattern2020serverless} & 2020 & \multicolumn{1}{c|}{\ding{55}} & \multicolumn{1}{c|}{\ding{55}} & \ding{55} & \multicolumn{1}{c|}{\ding{55}} & \multicolumn{1}{c|}{\ding{55}} & \multicolumn{1}{c|}{\ding{55}} & \multicolumn{1}{c|}{\ding{55}} & \ding{55} & \multicolumn{1}{c|}{\ding{55}} & \multicolumn{1}{c|}{\ding{51}} & \ding{55} & \ding{55} \\ \hline
% ROW 24 — white
\cite{challengeAppDev2021} & 2017 & \multicolumn{1}{c|}{\ding{55}} & \multicolumn{1}{c|}{\ding{55}} & \ding{55} & \multicolumn{1}{c|}{\ding{55}} & \multicolumn{1}{c|}{\ding{55}} & \multicolumn{1}{c|}{\ding{55}} & \multicolumn{1}{c|}{\ding{55}} & \ding{55} & \multicolumn{1}{c|}{\ding{55}} & \multicolumn{1}{c|}{\ding{51}} & \ding{55} & \ding{55} \\ \hline
% ROW 25 — grey (Our work — highlighted to stand out)
\rowcolor{rowgray}
Our work & 2025 & \multicolumn{1}{c|}{\ding{55}} & \multicolumn{1}{c|}{\ding{55}} & \ding{51} & \multicolumn{1}{c|}{\ding{55}} & \multicolumn{1}{c|}{\ding{55}} & \multicolumn{1}{c|}{\ding{55}} & \multicolumn{1}{c|}{\ding{55}} & \ding{55} & \multicolumn{1}{c|}{\ding{55}} & \multicolumn{1}{c|}{\ding{55}} & \ding{55} & \ding{55} \\ \hline
\end{longtable}
\vspace{0.5em}

\noindent
\textbf{Legend:} \\
\scriptsize
\begin{tabular}{ll}
\ding{51} & Covered or addressed in detail \\
\ding{55} & Not addressed or focused \\
\end{tabular}

\end{landscape}

% \\ What previous surveys have covered
\textbf{Related Surveys:} Existing surveys \cite{agbaje2022serverless}\cite{hassan2021survey}\cite{berkleyView2019}\cite{whatSCis2023}\cite{whatSCis2021} on serverless computing highlight the state-of-the-art features offered by the FaaS execution model, while other studies \cite{BaldiniChalOpp2017}\cite{hellerstein2018serverless}\cite{LannurienChalOpp2023}\cite{shafieiChalOpp2022}\cite{challengeAppDev2021} discuss unique challenges and opportunities in the application development and deployment process for FaaS. Furthermore, topics like resource management \cite{anupamaSurvey}\cite{resMan2023}, scheduling and autoscaling \cite{scheduling2018}\cite{schedAuto2023}, platform architecture \cite{shahradArchitecture2019}, cold start \cite{golec2024cold}, security \cite{archTechnicalPrimer2022} and performance evaluation of commercial as well as open-source platforms \cite{evalProd2018}\cite{evalMicroSvc2018} have attracted attention from researchers in the serverless computing domain. Complementary to these works, studies also address the FaaS workload characterisation \cite{pattern2020serverless}, community consensus on serverless applications \cite{eismannApp2022}\cite{latency2022}, and economic implications \cite{economy2024}\cite{economy2020} of the FaaS execution model. However, to the best of our knowledge, this is the first survey that focus on the function configuration aspect of the FaaS model, unravelling its importance and emerging state-of-the-art proposals in this direction. In this article, we identify certain aspects such as targeted workload characteristics, types of resources, deployment platform and key performance indicators to conduct a survey on function configuration techniques in FaaS. We propose a taxonomy of identified factors that influence function design, resource requirements, operational cost and performance guarantees of a function. This article further identifies gaps in the literature and highlights future research directions for right-sizing functions and enhancing the capabilities of FaaS environments to promote their increased adoption.

The rest of this article is organised as follows: Section \ref{sec:background} provides a brief background on serverless computing and introduces the Function-as-a-Service (FaaS) execution model. This gives an overview of the serverless characteristics, a high-level workflow from a developer and user perspective, existing FaaS platforms and their resource configuration offerings, leading to the motivation of this work. Section \ref{sec:taxonomy} presents the proposed taxonomy for function resource configuration and management and classifies existing studies. Section \ref{sec:futureres} comprises of the ideas for future research in this direction and Section \ref{sec:conclusions} concludes the study with an overview of our taxonomy. 

\section{Background}
\label{sec:background}

In this section, we present a brief overview of serverless computing, its popular compute service model, FaaS and its key features. Additionally, we discuss distinct characteristics of leading industry platforms and open-source FaaS frameworks to highlight their resource configuration strategy. Then, we introduce the scope of function resource configuration in serverless computing environments, its unique challenges and the motivation for our work.

\subsection{Serverless Computing}
\label{subsec:sc}

Over the years, cloud computing has enabled the shift from traditional ways of accessing IT resources to more available, affordable and scalable ways. But, with the emergence of micro-services and service-oriented architectures (SOA) \cite{hassan2021survey}\cite{berkleyView2019}, a new cloud-native service model of serverless computing came into existence. In this model, a cloud service provider takes charge of the server/resource management activities like resource planning, provisioning, allocation and scheduling. The idea of serverless computing builds on the resource abstraction provided by service models like Infrastructure-as-a-Service (IaaS) and Platform-as-a-Service (PaaS) to hide the complex resource management task on-demand while providing building blocks for application development. Thus, \textit{serverless} 
% does not mean the absence of servers to host and run the applications, but 
promotes the idea that developers need not worry about the underlying technology and focus on adding value to the businesses. Serverless offers seamless auto-scaling where resources are added or removed based on the demand and thus, users are charged according to the actual resource usage. The domain of serverless computing consists of a scalable function-based compute service, Function-as-a-Service (FaaS) and other complementary services like networking, message queue or storage under Backend-as-a-Service (BaaS) offerings.

\rev{Additionally, it is also important to note that the serverless paradigm has continued to evolve beyond the classical FaaS and BaaS model. Over the past several years, platforms such as Google Cloud Run, IBM Cloud Code Engine, and others have extended serverless principles, including scale-to-zero, pay-per-use billing, and event-driven execution to user-defined containers, longer runtimes, and more flexible resource configurations. This evolution, referred to by practitioners as \textit{Serverless 2.0} \cite{Femux2026}, decouples serverless benefits from the constraints of pre-defined function runtimes and enables a broader class of applications, including multi-container pods, multilingual workloads, and custom base images. Notably, Serverless 2.0 platforms such as Knative typically offer independent CPU and memory allocation and support higher degrees of per-instance concurrency. The present survey focuses on the classical FaaS execution model and its function resource configuration challenges, as this is where the body of configuration research is concentrated. Nevertheless, the principles and techniques surveyed here are increasingly applicable to the broader Serverless 2.0 context, and we identify this as an important direction for future configuration research.}

\subsubsection{Function-as-a-Service Execution Model}
\label{subsubsec:FaaS}

\begin{figure}
    \centering
    \includegraphics[width=1\linewidth]{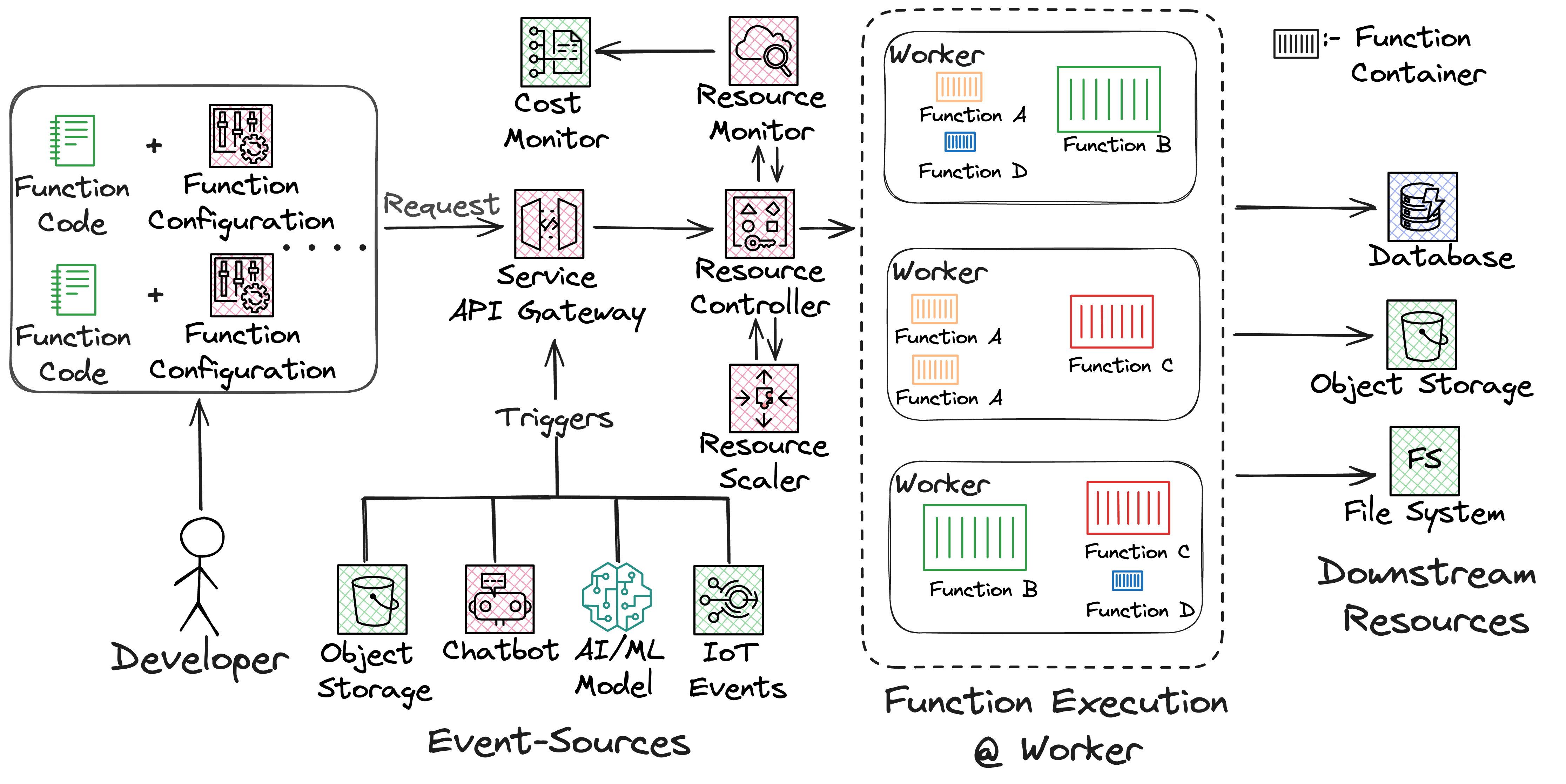}
    \Description{FaaS Workflow - Developer \& User Perspective}
    \caption{FaaS Workflow - Developer \& User Perspective}
    \label{fig:faas-workflow}
\end{figure}

FaaS was publicly introduced by AWS \cite{berkleyView2019} as a serverless compute service, AWS Lambda \cite{awsLambda}, and has since become a de facto for rapid application development, deployment and delivery. It offers an event-driven execution where applications are composed of independent functions and invoked via an event source or \textit{trigger}. A trigger could be an HTTP event, database or storage events or an IoT notification that executes the service and responds to an incoming request. 
% A \textit{function} is a code fragment or application logic that is the basic unit of development and deployment in FaaS. 
These functions usually serve a single purpose and are inherently stateless, highly scalable and meant to execute for a short period. Every function can be deployed as a container, lightweight Virtual Machine (VM) or a micro-VM \cite{Firecracker} and is associated with limited amount of compute resources. Functions are billed as per the resources used during execution and incur no idle costs. Furthermore, FaaS features scale-to-zero capability, where function resources are released after a period of inactivity. Currently, leading cloud providers such as Google, Microsoft, and IBM also offer FaaS platforms that support a wide range of applications, including event-driven websites \cite{ServerlessLand}, video streaming services \cite{ServerlessVideo}, multimedia processing \cite{NetflixAWS}, and CI/CD pipelines \cite{CapitalOne}. While the core function-based abstraction is similar among serverless platforms, they vary in their flexibility of resource allocation, billing granularity, and support for different function runtimes and BaaS services.
% This fine-grained pay-per-use pricing model and relief from resource management activities encourage developers to dedicate their efforts to adding business value rather than worrying about the underlying infrastructure. 
Figure \ref{fig:faas-workflow} depicts the FaaS workflow from a developer, user and service provider perspective. A developer uploads the function code along with the function configuration to execute, a user may interact with an event source to trigger these functions, and the service provider manages the platform and controls resource-related decisions.

\subsection{Key Characteristics of the FaaS Model}
\label{subsec:FaaSFeatures}

As per \cite{berkleyView2019}\cite{scSOTAli}, some of the key characteristics of current FaaS offerings can be summarised as follows - 

\paragraph{Event-driven: }FaaS is an event-driven offering where functions are invoked by various event sources such as HTTP requests, database changes, timers or orchestration events. The platform responds to these events in real-time by creating on-demand function instances. 
\paragraph{Hostless and elastic: } The FaaS execution model relieves users of complex resource management activities like server allocation, OS update and patching and reduces operational burden. Additionally, users are unaware of where the function is hosted, i.e., hostless, which allows for more flexible execution and pay-per-use pricing by the service provider. To support elasticity, FaaS offers seamless scalability where resources scale based on demand and are freed if there are no invocations.
\paragraph{Statelessness: } A serverless function executes within an isolated and temporary environment (container or a micro-VM) and does not maintain an in-memory state between invocations. This statelessness makes the function scale-out easier as the platform could instantiate parallel instances for concurrent requests without managing or transferring the function state between them.   
\paragraph{Short but variable execution time: } A typical serverless function lasts for a few milliseconds to \rev{minutes} \cite{shahradArchitecture2019}\cite{shahradServerlessWild2020}\cite{longTerm2023} and is effectively billed in units of milliseconds for actual running time. \rev{Production-scale characterisation studies have further revealed that execution durations and invocation patterns vary significantly across application types and exhibit long-term seasonal trends, underscoring the importance of workload-aware configuration strategies \cite{shahradServerlessWild2020}\cite{longTerm2023}\cite{Femux2026}. Platform-specific upper bounds also vary considerably, for instance, AWS Lambda permits execution of up to 15 minutes per invocation \cite{awsLambdaQuotas}.} However, the cold start time can significantly impact the perceived latency or performance of the functions that execute for very short durations. 
\paragraph{Memory and CPU Allocation: }Functions are often allocated operational resources and many FaaS platforms allow developers to specify the amount of allocated memory. Generally, runtime memory is a critical configuration that is exposed by the providers like AWS, Azure and Google where a function is allocated other resources like CPU and network bandwidth in proportion to memory. For example, AWS allocates compute power proportional to memory where 1 vCPU is allocated at 1769 MB of function memory. A function with more memory typically means more CPU power, leading to faster execution for compute-intensive applications. This also means that the function can process larger amounts of data, perform complex operations at speed and reduce overall latency. However, larger memory allocations also affect the operational cost, as the billing is typically done based on the running time and the allocated memory.
\paragraph{Execution timeout: } FaaS providers typically limit the execution time of a function before it is forcibly terminated by the platform. Developers must configure their function with a long enough function timeout to finish its task. For tasks that execute for a few milliseconds or seconds, setting a short timeout is suitable, whereas a longer running task such as multimedia processing \cite{ServerlessVideo}, a longer timeout is intended. Providers such as AWS allow timeouts up to 15 minutes, whereas Azure Functions may provide longer timeouts, i.e., more than 30 minutes, based on the subscribed plan. The timeout is also significant as it prevents a function from getting stuck in an infinite loop and saves developers from exorbitant costs. This way, a function is automatically stopped after exceeding the timeout.
\paragraph{Ephemeral Storage: }In addition to the compute resources, a function also has a temporary configurable storage attached to it. \rev{The ephemeral storage is typically only available during the life of a function instance} and is useful in applications that require a file system during execution for processing or intermediate data storage, like file download. 
\paragraph{Concurrency: } Function concurrency refers to the number of simultaneous incoming requests a function can handle. FaaS platforms typically react to the incoming demand and may start multiple function instances concurrently. However, to reduce the effect of cold start and improve function performance, developers can also configure provisioned concurrency \cite{awsLambda} or minimum function instances \cite{runFunctions} that keep a certain number of function instances warm and ready to execute. This is critical for latency-sensitive applications, but it does incur some extra costs for keeping the resources warm and idle.

\subsection{Function-as-a-Service Platforms and Frameworks}
\label{subsec:platforms}
Currently, many CSPs like AWS, Microsoft Azure and Google offer various services under the serverless umbrella, including compute (FaaS), database, networking and storage. Among the popular FaaS platforms, AWS Lambda \cite{awsLambda} covers over 50\% of market share \cite{eismannApp2022} followed by Azure Functions \cite{azureFunctions} and \rev{Cloud Run Functions \cite{runFunctions} (formerly Google Cloud Functions \cite{gcf})}. These proprietary solutions require functions to be developed and integrated with their respective ecosystems, leading to vendor lock-ins. To counter this, open-source serverless frameworks have emerged. Frameworks like OpenFaaS \cite{openfaas}, Apache OpenWhisk \cite{openwhisk} and Knative \cite{knative} are predominant as they support public, private, and hybrid cloud deployments,  introducing a level of flexibility to serverless offerings.

\subsubsection{AWS Lambda}
\label{subsubsec:AWS}
\rev{AWS Lambda is an} event-driven compute service that supports various language runtimes such as Python, Node.js and Rust\rev{, as well as custom runtimes via Docker container images, enabling developers to bring their own runtime (BYOR) and execute functions in virtually any language or environment \cite{awsLambdaQuotas}}. While it integrates with various AWS and external services and event triggers like Amazon S3 and Apache Kafka, it offers an important aspect of configurable resources to developers. Lambda is subject to certain quotas and limits that developers must explore to optimise the operational cost and performance of the function. The most important consideration is a function's memory allocation, which developers can configure between 128 MB and 10,240 MB in 1 MB increments, proportionally allocating other resources like CPU power. For example, 1769 MB of memory allocates an equivalent of 1 vCPU to the function. While increasing the memory gives more CPU share and speeds up a function execution for a compute-intensive task, it may lead to increased operational cost per invocation. A Lambda function is allowed to run for a maximum duration of 15 minutes, and developers must set an appropriate \textit{timeout} to prevent any unexpected costs. Additionally, functions may be configured with ephemeral storage of up to 10,240 MB for tasks that may require intermediate file download or processing. AWS also puts a limit on the size of data passed to a function, i.e., the \textit{payload}, specifically, 6 MB for synchronous request-response, 200 MB for streamed response, and 256 KB for asynchronous invocations \cite{awsLambdaQuotas}. As these functions scale in response to the demand, Lambda functions can scale up to 1000 unique execution environments every 10 seconds to handle the concurrent invocations. \rev{For asynchronous invocations, Lambda automatically retries failed events for up to 6 hours, enabling resilient handling of workloads that temporarily exceed the concurrent execution quota of 1000 instances \cite{awsLambdaQuotas}.} Apart from the configurable settings, Lambda does offer generous free-tier usage limits \cite{awsLambdaPricing}, however, understanding the effects of resource configuration is crucial for developing cost-effective and performant functions. Lambda further integrates with services like AWS Step Functions \cite{awsStepFunctions} to support workflow application scenarios, where respective solution performance overheads and pricing are applicable in addition to Lambda invocation.

\subsubsection{Microsoft Azure Functions}
\label{subsubsec:Azure}
Microsoft announced the availability of its Azure Functions in November 2016, where an application-first approach is taken for development to represent a group of functions as an application. Azure Functions supports multiple language runtimes, including C\#, Java, PowerShell and Python \cite{azureFunctions} while integrating with other Microsoft and external services event triggers, including Azure Cosmos DB events and message queues. Unlike other FaaS platforms, a defining characteristic of Azure Functions is its flexible hosting and resource configuration schemes \cite{AzureConsumptionPlan} that give developers control over the application performance and cost. It offers different hosting options, such as premium, consumption or dedicated, which define how and what type of resources are supported, configured and billed. The most common option is the consumption plan \cite{AzureConsumptionPlan} that offers a pay-as-you-go resource consumption and billing scheme where the underlying host will be allocated up to 1.5 GB of memory with a variable CPU share. For more complex workloads, schemes like the \textit{flex consumption plan} are also offered, where instances with memory allocation up to 4 GB and proportional CPU share can be selected. While configuring more memory gives more processing power, Azure recommends using a 2048 MB memory instance out of available options for most scenarios. 
% out of 512 MB, 2048 MB, and 4096 MB based on the application needs. 
Azure allows a unique configuration of the maximum duration of a function, where a timeout of 10 minutes can be configured in a consumption plan, while no timeouts are enforced in a flex consumption plan, with a default capped at 30 minutes. Azure further offers both ephemeral and persistent storage as part of its hosting plans where temporary storage could range from 0.5 GB to 140 GB per instance and persistent storage can be configured between 1 GB and 1000 GB per hosting plan. In addition to this, functions could scale to as few as 100 instances in a dedicated plan, up to 1000 instances in a premium plan, for a more aggressive scaling with a maximum allowed request \textit{payload} size of 210 MB. Azure further allows development of stateful applications and workflows in addition to simple stateless functions via Durable Functions \cite{azureDurableFunction}. These features, combined with various resource configuration options, allow developers to explore settings and fine-tune the application performance.

\subsubsection{Cloud Run Functions}
\label{subsubsec:GCF}

Rebranded in 2024 from \textit{Cloud Functions Gen 2} \cite{runFunctions}, \textit{Cloud Run Functions} unify previous function deployments (Cloud Functions Gen 1 and Gen 2) under Cloud Run's fully managed container platform \cite{runFunctions}. Cloud Run Functions support several programming languages, including Python, Java, Go, Node.js, .NET, PHP, and Ruby, where the runtime base image and container lifecycle are fully managed by Google. They support various \textit{triggers} such as HTTP requests, Pub/Sub messages, Cloud Storage changes, and Firestore database events via Eventarc integration. Memory allocation is configurable up to 32 GiB with flexible CPU allocation proportional to memory, ranging from fractional vCPUs during idle phases to up to 8 vCPUs during active request processing \cite{gcfCPUAllocation}. By default, CPU is allocated only while processing requests and during container startup and shutdown but can be configured to run continuously if needed. Request timeouts default to 5 minutes but can be extended up to 60 minutes, accommodating long-running workloads. Cloud Run Functions support concurrency of up to 1000 simultaneous requests per instance, improving resource efficiency and reducing cold start latency. 
% Instances recycle based on platform policies to maintain freshness. 
These functions seamlessly integrate with the wider Google Cloud ecosystem and support complex workflows through the Workflows service \cite{googleWorkflows}, combining the scale and flexibility of container-based deployment with the simplicity of function-as-a-service.

\subsubsection{Knative}
\label{subsubsec:Knative}

It is a Google-sponsored Kubernetes-native platform that supports serverless workloads. Knative \cite{knative} leverages the underlying Kubernetes primitives of container orchestration, scaling, scheduling and configuration to provide tools that automate the task of continuous integration and continuous delivery (CI/CD).  
% The two main components, Knative Eventing and Knative Serving, work together to manage tasks and applications. 
Knative deploys functions as Kubernetes \textit{pods}, supporting language runtimes such as Python, Rust, Go and TypeScript. Functions can be invoked via \textit{triggers} such as HTTP events and events that conform to CloudEvents. Knative offers a scale-to-zero capability that supports concurrency and request-per-second metrics for autoscaling. The platform also provides fine-grained control to allocate resources like memory and CPU to functions. Kubernetes primitive of resource requests and limits can be leveraged for specifying these values. Another key configuration aspect is concurrency, which defines the number of concurrent requests a single service instance can handle. By default, Knative often sets a concurrency limit of 100 concurrent requests per instance, but this can be adjusted. Setting a higher concurrency can be cost-effective as it reduces the number of running instances, but it requires careful tuning of CPU and memory to ensure each instance can handle the increased load without performance degradation. \rev{A recent large-scale production characterisation of a Knative-based platform \cite{Femux2026} reveals important empirical insights into how users configure the concurrency parameter in practice. It reports that the majority of applications are deployed with non-zero minimum scale settings, indicating that developers prioritise latency over cost efficiency, and that concurrency limits vary widely across deployments. These findings have direct implications for configuration strategies that assume default or uniform concurrency settings.} Additionally, Knative integrates well with different monitoring and logging solutions like Prometheus and Grafana, FluentBit or ElasticSearch to collect several metrics from platform components.
% such as autoscaler, serving and controller.  

\subsubsection{OpenFaaS}
\label{subsubsec:Openfaas}

OpenFaaS is an open-source Kubernetes-native serverless platform that simplifies function development, deployment and management. Functions can be deployed as Open Container Initiative (OCI) images with toolkits like Docker and developed in runtimes such as Python, Ruby, Go and Java. OpenFaaS \cite{openfaas} offers subscription-based access like community, pro and enterprise editions with different levels of service offering. It supports a number of function triggers such as HTTP requests, Cron jobs, AWS events and Postgres database events. Additionally, functions can be configured with three autoscaling types of request-per-second, capacity and CPU within a range of 1 to 5 instances (community edition), and a scale-to-zero provision is available with the Pro edition to release resources after an idle duration of 15 minutes \cite{openfaas}. The CPU and memory resource requirements and limits can be configured for a function as a Kubernetes primitive and limit the concurrent requests executing inside a container or the timeout by setting an environment variable. Furthermore, OpenFaaS can be set up on different environments such as AWS EKS, Azure AKS, Google GKE, etc.

\subsubsection{Apache OpenWhisk}
\label{subsubsec:openwhisk}
IBM introduced OpenWhisk \cite{openwhisk} as an \textit{open}-source serverless platform in February 2016 with an idea to quickly run users' code and \textit{whisk} or release its resources. It runs functions in response to different events from various sources such as mobile and web applications, databases, scheduled jobs and sensors. Its architecture is powered by multiple open-source technologies such as Docker, Apache Kafka, CouchDB and Nginx engine. The OpenWhisk programming model supports code written in Java, Python, .NET, PHP and Swift\rev{, while also supporting custom runtimes beyond the standard set via Docker-based deployment \cite{openwhiskDetails}}. Functions are automatically scaled and can be configured with a number of options to limit the system usage \cite{openwhiskDetails}. A function can execute up to 5 minutes per invocation with a concurrent invocation rate of 120 requests per minute. Functions can be allocated memory between 128 MB and 512 MB, have a maximum payload size of 1 MB, a maximum code size of 48 MB, and can be configured with a per-action maximum function concurrency.

\subsection{Resource Configuration in FaaS Model}
\label{subsec:resconfig}

FaaS execution brings ease to modern application development and deployment by allowing developers to focus on logic rather than resource governance tasks. However, the significant influence of varied configuration settings on a function's performance and operating cost introduces a unique set of bottlenecks, including a trade-off between performance guarantees and developer complexity. Therefore, finding the right balance or combination of available resource settings is crucial in a constrained FaaS environment.

\subsubsection{Motivation}
\label{subsubsec:Motivation}

% Intro to the problem
In FaaS, developers are usually responsible for configuring the desired amount of resources for their functions required during execution. Serverless platforms like AWS Lambda \cite{awsLambda}, Microsoft Azure \cite{azureFunctions}, and Cloud Run Functions \cite{runFunctions} generally tie resources like CPU share and network bandwidth to the allocated memory, leaving developers with limited visibility of resources. Additionally, the run-time performance of a function varies significantly with the allocated resources and may result in unexpected execution cost \cite{Sizeless}. Therefore, a developer must make the optimal configuration choice in order to reduce the operating cost while guaranteeing successful execution. This leads to a complex resource selection and allocation optimisation challenge that could result in wasted or insufficient resources for the execution. For example, 
% Explain wastage/performance aspect 
allocating too many resources (more memory and CPU shares) than required for successful execution results in resource over-provisioning, faster and expensive invocations. On the other hand, configuring fewer than required resources or under-provisioning may lead to failed invocations, increased latency and poor user experience. Therefore, striking the right balance between performance, user-perceived latency, and cost is difficult. 

In addition to this, a function's resource requirement may change based on the workload, request parameters, input size and its characteristics, etc., and static configurations may not suffice or be inefficient for all incoming invocations \cite{UCCSiddharth}\cite{ParrotFish}. For instance, a resource setting for a small-sized input data may not be adequate for a large one, and manually adjusting the resources for every potential input is infeasible. Thus, determining the optimal resource requirement of a function for different workloads and use cases is a tedious task and often involves extensive monitoring and experimentation. The problem is further exacerbated for applications involving multiple functions, where the impact on a single function's performance can have a cascading effect. However, optimising a single function in isolation for performance may not guarantee an end-to-end performance improvement. Apart from the workload characteristics, the performance of a function is also degraded due to the inherent function cold start. This is an actively researched challenge \cite{golec2024cold} of FaaS platforms and studies \cite{LambdaColdStart}\cite{LambdaColdStartYanCui}\cite{LambdaColdStartPluralSite} suggest that higher memory or resource configuration reduces function cold start time significantly, in addition to other contributing factors such as language runtime and deployed code size. In order to mitigate this overhead, CSPs usually offer provisioned concurrency or a minimum function instances setting that allows developers to maintain a pool of warm instances to reduce the cold start impact on performance. However, this requires developers to balance anticipated demand against the costs of idle resources, making resource configuration an even more complex task. Although cold starts do not affect the actual execution time of the function (i.e., the time the code is actively running once initialised), the selection of an optimal timeout is crucial, both for providing sufficient time for the function to complete its task and for preventing unnecessary costs due to runaway or stalled executions. The timeout serves as a safeguard, ensuring that a function doesn't run indefinitely, which could lead to exorbitant and unforeseen charges. But, as a function's execution time can vary significantly based on the workload, setting a static timeout may result in failed invocations for functions with longer-running tasks. The challenge of static configuration also extends to the temporary storage, where functions leverage it for tasks like file processing or dataset download, and functions may fail if they exceed the limited, non-persistent storage. This forces developers to configure larger and more expensive configurations or introduce complex external storage solutions like \rev{Amazon EFS (for POSIX-compliant file system access) or} Google Cloud Storage or Amazon S3 \rev{ (for object storage)}, to manage function data and state, further complicating the configuration process.

Prior research studies \cite{WithGreatFreedomComesGreatOpportunity}\cite{ParrotFish} have shown that parameters such as execution time and cost vary non-linearly with resource configuration, where factors like input characteristics, such as input size, workload and resource requirements, impact function performance. To address the fluctuating performance, developers generally profile their functions for the anticipated workload and allocate a static resource configuration strategy. This profiling is not only tedious but also provides a rough estimate of a function's resource requirements, generally resulting in a sub-optimal resource allocation. Furthermore, offering efficient resource configuration solutions enables developers to fully offload function-related decisions to CSPs, thereby experiencing an end-to-end serverless workflow. Therefore, we identify function resource configuration as a fundamental challenge in the serverless computing resource management paradigm and classify different configuration-related elements by reviewing the existing literature.

\begin{figure}
    \centering
    \includegraphics[width=1\linewidth]{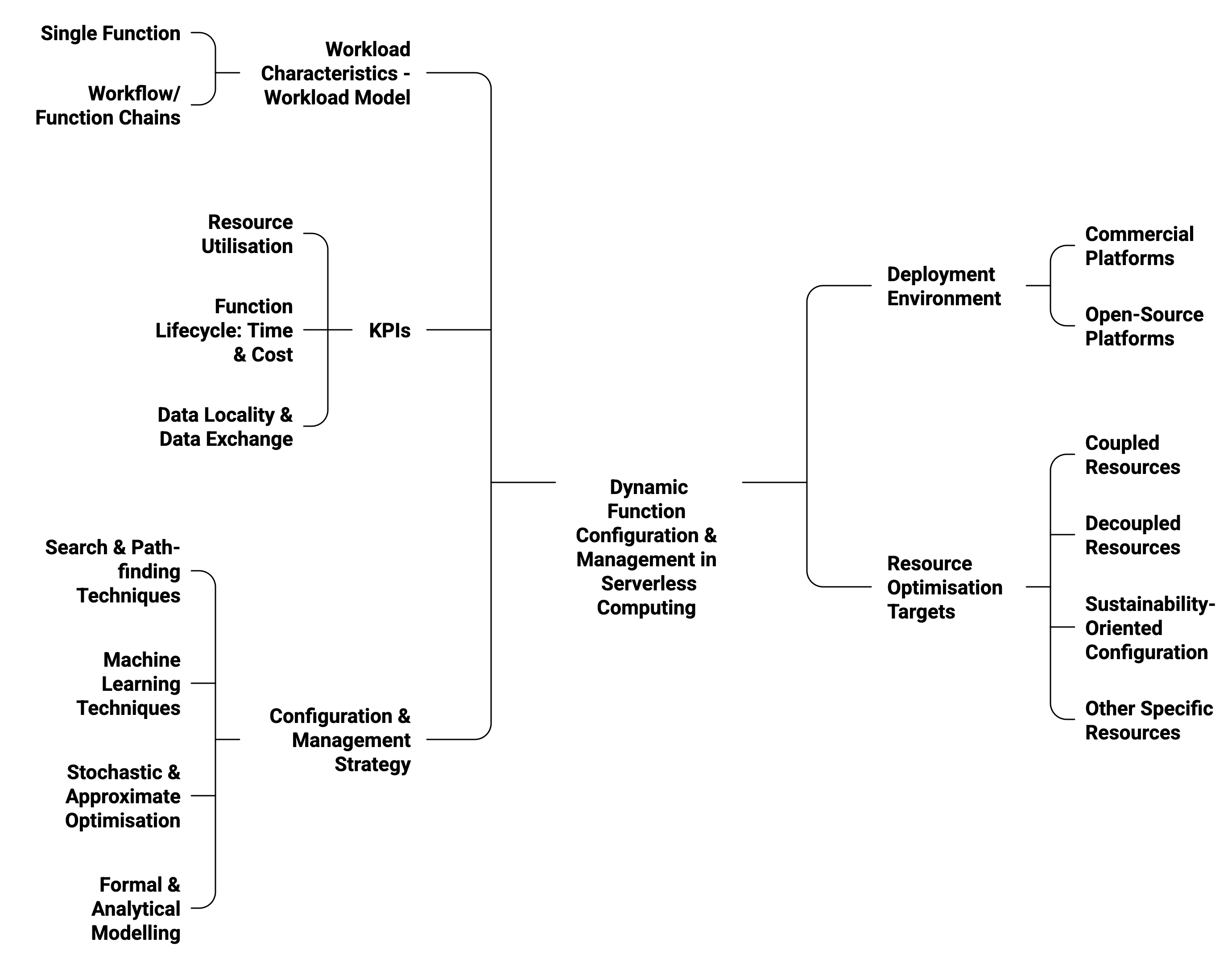}
    \Description{Function Configuration Taxonomy}
    \caption{A Taxonomy of Dynamic Function Configuration \& Management in Serverless Computing}
    \label{fig:full_taxonomy}
\end{figure}

\section{The Taxonomy}
\label{sec:taxonomy}

This section outlines the identified taxonomy of the related literature for dynamic function configuration. The high-level categorisation, Figure \ref{fig:full_taxonomy}, is comprised of the key elements of function configuration strategy, factors affecting configuration decisions such as \rev{Key Performance Indicators (KPIs)}, optimisation goals, workload characteristics and deployment environments. The proposed taxonomy is highlighted in Figure \ref{fig:full_taxonomy} and a more granular discussion is presented in the following sections.

\rev{The discussed five dimensions of the taxonomy are not independent as they form a causal hierarchy in which the \textit{workload characteristics} of a serverless application (Section~\ref{subsec:workloadCharacterisation}) determine what configuration decisions are relevant and what performance targets are meaningful. These targets are captured as \textit{KPIs} (Section~\ref{subsec:kpi}), which in turn drive the selection of \textit{resource optimisation targets} (Section~\ref{subsec:optimisedResource}). The \textit{deployment environment} (Section~\ref{subsec:deploymentEnv}) acts as a structural constraint: commercial platforms restrict developers to coupled resource models, whereas open-source frameworks permit decoupled configuration, making the deployment environment a causal factor that determines which optimisation targets are technically feasible. Finally, the \textit{configuration and management strategy} (Section~\ref{subsec:strategy}) captures the algorithmic or modelling approach used to navigate the resulting configuration space. This hierarchical structure from workload characteristics, through KPIs and deployment constraints, to optimisation targets and strategy reflects the decision-making process that practitioners and researchers follow when addressing function resource configuration in serverless environments.} 

\rev{Moreover, the selection of these five dimensions is grounded in the existing literature where each dimension captures a class of factors that independently governs configuration decisions and cannot be fully subsumed by another. For example, factors such as input data characteristics, payload size, and invocation patterns are treated as sub-characteristics within the workload characteristics rather than as standalone dimensions, because they operate at the level of individual function inputs and shape the workload model rather than structurally constraining which resources can be configured or which optimisation strategies are applicable.}

\subsection{Workload Characteristics}
\label{subsec:workloadCharacterisation}

This section discusses the existing literature under the scope of workload characteristics that have a direct impact on function resource configuration. The identified elements of the workload model and nature form the basis of different configuration approaches, selection of optimisation goals and key performance indicators recognised in the explored works.

\subsubsection{Workload Model}
\label{subsubsec:workloadModel}

Workload model refers to the composition of serverless applications using one or more functions as components with their respective resource requirements. FaaS has been proven suitable for a diverse range of use-cases; however, the application's expected quality-of-service (QoS) and the number of functions in an application, also known as function cardinality, are mutually dependent. A function's performance metrics, execution time and run-time cost are usually proportional to the allocated resources, while the application's end-to-end performance fluctuates with various function configurations. Furthermore, as the function cardinality grows, an application's performance-cost trade-off becomes non-trivial. Therefore, considering an application workload model is a fundamental factor for tuning function resources.

\paragraph{Workflow/Function Chain: }A function chain or a workflow is a series of functions that often execute in specific sequence, order or in parallel. The output of one function becomes the input of another, enabling composition of complex, multi-step applications \cite{Wisefuse}\cite{whatSCis2023}. These function chains are critical for scenarios where a single, short-lived function can not handle the complex task such as data processing and ML workflows. Generally, orchestration tools such as AWS Step Functions \cite{awsStepFunctions} and Azure Durable function orchestrations \cite{azureDurableFunction} are leveraged to build, manage, and execute workflows. Each function or step in the workflow requires individual settings such as timeout, concurrency and resource configurations where the overall performance and execution of the workflow is governed by these individual units. Consequently, when functions in a chain are executed, configuration of downstream services is also vital to avoid overwhelming them or exceeding any platform-specific limits. While individual function executions are billed per-execution, the chains are typically charged based on number of state transitions. Therefore, efficiently designing the workflows with minimal steps and executions can result in reduced overall operational costs.    
Elgamal et al. \cite{Costless} introduce a possibility of function fusion, i.e., combining multiple functions together and configuring their memory to reduce the execution cost of function workflows. 
% The researchers posit that workflow costs are based on number of state transitions where function fusion and their memory adjustment are the ways to reduce the impact on SLOs. 
In another work \cite{PredictingTheCostsOfWorkflows}, Eismann et al. criticise the static resource allocation by CSPs for individual function invocations and stress the exacerbated response times of function chains due to inter-function dependency. 
% To this end, they propose a Mixture Density Network (MDN) based input/output parameter to response time relationship modelling and Monte-Carlo simulation to estimate workflow costs. 
Lin and Khazaei \cite{ModellingandOptimisationOfPerformanceCost} presented analytical models to predict end-to-end execution time and cost of workflows as Directed Acyclic Graph (DAG) and find optimal function configuration under performance constraint. 
% These models convert a function chain in probabilistic DAG and apply a Probability Refined Critical Path Greedy (PRCP) heuristic to find an optimal function configuration under a performance or a budget constraint. 
Likewise, Wang et al. \cite{EvolutionaryProbabilistic} present a Genetic Algorithm-based approach for memory configuration of serverless workflows owing to the lack of platform transparency for users. Bhasi et al. \cite{Kraken} consider latency-critical dynamic DAGs to minimise the function resources, particularly the spawned containers, by tuning the number of requests served by a container without violating the SLO. A multi-stage configuration framework to optimise resources at each step of the function workflow is introduced by Wen et al. \cite{StepConf}. The authors bring attention to proper resource configuration in a function chain to meet SLO and save costs. Comparable to previous works, \cite{SLAM} identifies a research gap for function chain or workflow configuration and discusses an SLO-aware configuration tuning tool to improve the overall throughput of the application.

Zhang et al. \cite{VideoMeasurementStudy} perform a measurement study to identify the impact of user-controllable knobs on function-based video processing applications from a developer's perspective. They share that memory size, platform selection and underlying infrastructure have an impact on application performance. The research works \cite{Llama}\cite{EdgeAssistedVideoAnalytics}\cite{charmseeker} explore FaaS for auto-tuning of video analytics and processing pipelines as DAGs to meet diverse latency targets while minimising runtime costs. The authors identify the inter-function input/output relationship, the effect of video content and pipeline-specific factors on cost and performance, and independent function resource requirements to optimally tune the configuration of involved functions. Similarly, Orion \cite{Orion} aims to meet probabilistic guarantees for end-to-end execution time in function DAGs such as video analytics and ML pipelines by right-sizing individual functions in addition to co-locating and pre-warming them. Aquatope \cite{Aquatope} emphasises satisfying QoS constraints of a multi-stage serverless workflow and reducing resource wastage by optimally configuring functions of a workflow. The research outlines an inter-dependency of function cold start and resource configuration, where different amounts of resources are required to achieve the same QoS guarantees. In addition, \cite{CEScaling} targets serverless ML workflows to provide an efficient fine-grained resource allocation and partitioning strategy to distribute the resources among the different stages of the workflow and reduce wastage. Furthermore, \cite{Astrea} finds resource allocation and configuration in a data-intensive analytics application challenging and proposes a Graph theory-based solution in Map-Reduce (MR) style function-based applications.

\paragraph{Single-function application: }A large body of existing literature deals with the resource configuration of single-function applications that perform a dedicated task. In FaaS, a function is a fundamental unit of deployment and the lowest level of available customisation interface. Because functions are typically designed to perform intricate, individual tasks, researchers have concentrated their efforts on optimising the resources for single-function applications to ultimately improve overall workflow performance. However, these studies often overlook the complexities of managing and optimising the resource for inter-connected and dependent function within a larger workflow context. Kim et al. \cite{CPUCap} investigate the resource allocation for functions in multi-tenant serverless platforms and introduce a dynamic CPU allocation resource manager to minimise the resource contention among serverless functions. The study in \cite{COSE.V1} explores the statistical relationship between function memory configuration and its execution cost and time to optimally configure function memory that meets user-defined constraints. The authors posit that this problem is further exacerbated for function workflows. Suresh et al. \cite{Ensure} take a different approach towards CSP-focused function-level resource management on bare metal servers to regulate CPU shares of co-located functions and improve resource utilisation while satisfying application SLO. Tang and Yang \cite{Lambdata} criticise treating each function as a black box and discuss the opportunities of data caching and function co-location. They introduce explicit read/write data intents for serverless functions by developers to perform execution speed and runtime cost optimisations.

\rev{Mahmoudi and Khazaei} \cite{PerformanceModellingSC} present an analytical performance model to calculate essential performance metrics of individual functions in a steady-state that can be used to tune resource configurations. In another work \cite{RDOF}, Zhu et al. suggest an Integer non-linear programming problem to address memory configuration that is helpful in function deployment. Spillner discusses dynamic auto-tuning of serverless functions in \cite{RMFunctionsProfiling} and emphasises the importance of predictions, where functions can utilise freed-up resources from others to reduce resource wastage. Similar to this idea, \cite{OFC} proposes an opportunistic function cache that configures idle and extra invoker memory to resize the cache for data locality-aware function executions. The framework determines the right amount of function memory in order to vertically scale the invoker node cache memory for storing data objects. While research works like \cite{UCCSiddharth}\cite{Cypress}\cite{CPUTAMS}\cite{Sizeless} extensively focus on resource configuration via prediction-based input-size aware function memory configuration, input-sensitive request scheduling or dynamic CPU-share allocation, \cite{ParrotFish}\cite{OptimizingMemoryAllocationinaServerless} concentrate their efforts on memory allocation optimised function scheduling and an online parametric regression-based function memory configuration. These studies aim at function-level resource configuration decisions to improve utilisation, and reduce execution time and costs while satisfying user SLOs.

\rev{\textbf{Research Gaps:} While the surveyed works address both single-function and workflow-level configuration, several gaps remain. First, the majority of workload characterisation studies rely on older production traces \cite{shahradArchitecture2019} or synthetic benchmarks; more recent large-scale traces from production open-source platforms reveal that workload patterns, concurrency configurations, and cold start behaviour are evolving in ways not captured by earlier models \cite{shahradServerlessWild2020} \cite{longTerm2023}. Second, input-sensitive configuration research predominantly targets memory, with limited treatment of how varying input characteristics interact with timeout and concurrency settings simultaneously. Third, cross-function dependency in workflows remains underexplored from a configuration standpoint where most workflow studies optimise functions independently rather than jointly under shared resource budgets. These gaps motivate the future directions discussed in Section~\ref{sec:futureres}.}

\subsection{Deployment Environment}
\label{subsec:deploymentEnv}

This section analyses the existing literature for the type of serverless environment used for experimental setup and evaluation. In addition to commercially available FaaS platforms that expose limited configurable parameters, a number of open-source frameworks 
% and simulation environments 
are available to acquire a deeper understanding of FaaS architecture. Therefore, a discussion of these deployment environments is important to recognise and appreciate the feasibility of numerous research works in serverless systems. \rev{Critically, the deployment environment is not merely contextual background, it is a causal determinant of what resource configuration strategies are technically feasible. Commercial platforms such as AWS Lambda and Azure Functions couple CPU allocation to memory, constraining researchers and developers to coupled resource optimisation. Open-source frameworks such as OpenFaaS, Knative, and Apache OpenWhisk expose independent CPU and memory controls, enabling decoupled configuration studies. This structural difference directly governs which resource optimisation targets (Section~\ref{subsec:optimisedResource}) are accessible for a given study, justifying its placement as a distinct dimension in the taxonomy.}

% \textbf{\textcolor{red}{pie chart of distribution}}

\subsubsection{Commercial Serverless Platforms}
\label{subsubsec:commercial}

Commercial FaaS platforms such as AWS Lambda \cite{awsLambda} and Microsoft Azure \cite{azureFunctions} are widely utilised for conducting research owing to their increased adoption and real-world relevance. These platforms, however, offer limited architectural visibility and configurable function parameters due to their proprietary nature, forcing developers and researchers to infer their system behaviour through experimentation. Consequently, a number of research works aim to address these limitations by focusing on distinct function configuration and optimisation opportunities for performance enhancement and operational cost savings. As indicated in section \ref{subsec:platforms}, AWS dominates the FaaS cloud market among CSPs, with several studies utilising AWS Lambda for their research proposals and experimental setups. The study \cite{Costless} leverages applications from AWS Serverless Application Repository for comprehensive function performance profiling. The researchers explore function fusion on AWS Lambda, where a sequence of functions is combined. This technique, along with splitting functions between edge and cloud resources and allocating memory, is used to find a solution that minimises the application's price while keeping the latency under a certain threshold. The decision to fuse functions along with memory allocation is non-trivial as it has implications on both operational cost and execution latency. Akhtar et al. \cite{COSE.V2} conduct experiments on single function applications that focus on CPU, memory, network, and I/O-intensive tasks spanning across edge and cloud environments on AWS. They use performance monitoring and statistical learning based method to predict the execution time and cost of a serverless function at unseen configurations. This helps them to select the optimal memory configuration and function placement strategy that minimises not only the runtime costs but also meet user defined performance thresholds.

Similarly, Eismann et al. \cite{Sizeless} leverage Lambda to collect application monitoring data to build a large-scale synthetic dataset for ML-driven memory optimisation. The authors construct a multi-target regression model to predict execution time of a function to recommend the optimal memory configurations. On the other hand, Zubko et al. \cite{MAFF} present a function memory optimisation framework built atop AWS services. It utilises algorithms such as binary search and gradient descent to find the best function memory setting for a given cost or execution time objective. A recent work by Moghimi et al. \cite{ParrotFish} design a recommendation framework that can tune and refine resource allocation for Lambda functions. The framework employs online parametric regression technique to find and recommend optimised memory configuration while reducing the exploration cost. Moreover, the researchers in \cite{StepConf} employ Lambda alongside AWS Step Functions to enhance workflow configuration. Also, the study \cite{Astrea} implements map-reduce (MR)-style applications on AWS Lambda to efficiently allocate resources across pipeline functions. The research \cite{ModellingandOptimisationOfPerformanceCost}\cite{PerformanceModellingSC} analyses and represents the performance-cost trade-off of serverless platforms such as Lambda to refine function memory configurations. Accordingly, Xu and Lloyd \cite{CPUTAMS} propose a workload-agnostic memory selection method utilising regression modelling and perform experiments on commercial cloud platforms by AWS, IBM, Microsoft Azure and Google. Likewise, Eismann et al. \cite{PredictingTheCostsOfWorkflows} predict the workflow costs on GCF by leveraging performance monitoring data via ML and simulation techniques. Similarly, Agarwal et al. \cite{UCCSiddharth} leverage AWS Lambda and related serverless services to propose a multi-output regression model that optimally configures functions based on their input size. A number of studies \cite{Llama}\cite{charmseeker}\cite{VideoMeasurementStudy} have taken advantage of serverless platforms for video processing and analytics applications where \cite{VideoMeasurementStudy} discusses the impact of resource configuration on function performance, while \cite{Llama}\cite{charmseeker} direct their proposals for efficient pipeline resource configuration on AWS Lambda and Google Cloud function services. 

\subsubsection{Open-Source Serverless Frameworks}
\label{subsubsec:open-source}

In addition to large-scale commercial serverless services, many open-source serverless frameworks have also been introduced, such as OpenFaaS, Apache OpenWhisk, and Knative. Unlike commercial offerings, where function resources such as CPU and network bandwidth are allocated either proportional to memory or are selected from a pre-set list of configurations, open-source frameworks offer greater flexibility in configuring function resources. Therefore, these platforms have also been widely used for recording FaaS experiments and exploring different research opportunities. 

Tang and Yang \cite{Lambdata} in their work leverage OpenWhisk controller to make function input and output data locations explicit to optimise it for caching, scheduling or co-location. Similarly, Mvondo et al. \cite{OFC} integrate an ML-based agent with an OpenWhisk controller to configure function memory and reclaim any extra or unused memory for offering a cost-effective cache system. A short study by Pandey and Kwon \cite{OptimizingMemoryAllocationinaServerless} emphasises memory over-allocation and data locality of several functions by optimising scheduling decisions on OpenWhisk. Another research \cite{Lachesis} introduces an online function right-sizing method that is integrated with OpenWhisk components to allocate CPU resources to functions. Unlike the previous works, Zhuo, Zhang and Delimitrou \cite{Aquatope} focus on multi-stage application resource over-provisioning, where they introduce a custom controller and manager on the OpenWhisk platform to configure functions. 

On the other hand, Mahmoudi and Khazaei \cite{PerformanceModellingSC} develop an analytical model inspired by the Knative framework to predict performance metrics like response time and replica count and leverage it for resource configuration decisions. On a similar note, \cite{Latest} utilises a feedback loop controller in conjunction with the Knative platform to estimate function resources and applies the corrective values based on a performance threshold. A research by Cordingly et al. \cite{CPUTAMS} experiment on self-hosted OpenFaaS clusters to determine the efficacy of CPU time accounting principles and leverage its effect on execution time and cost for function resource configuration. Taking a different approach to input-size dependent performance of functions, Bhasi et al. \cite{Cypress} devise a resource management framework on top of OpenFaaS that batch and reorder incoming requests to fulfil agreed SLOs and minimise resource wastage. Alternatively, the research \cite{FireFace} exploits application-specific function features to estimate function sizes to deploy them on an edge network using an OpenFaaS cluster. Furthermore, Bilal et al. \cite{WithGreatFreedomComesGreatOpportunity} present resource decoupling for function auto-configuration and verify different existing approaches on an OpenFaaS cluster to suggest function resource allocation and adjustment as an opportunity to explore. In addition to well-known FaaS frameworks, including the ones running on top of Kubernetes such as OpenFaaS and Knative, a distributed programming framework like Dask \cite{Dask} has also been explored for function resource configuration and proposing a process-based serverless execution and resource cap control mechanism. Another study \cite{Costless} jointly uses a commercial serverless platform with a Raspberry Pi-based environment to emulate the edge locations for the function configuration and placement problem.

\subsection{Key Performance Indicators}
\label{subsec:kpi}

In this section, we analyse the key objectives identified in the existing literature that drive research on resource configuration in FaaS. These objectives are related to specific performance metrics that define the level of abstraction targeted by individual studies. A function's configuration, such as memory, CPU, or data locality, has a direct or indirect impact on these key indicators and reflects the quality of service experienced by the user or the performance guarantees offered by the CSPs. Therefore, performance metrics such as throughput, execution time and cost, resource utilisation, and data locality are emphasised in the existing studies while optimising the resource allocation.

\subsubsection{Resource Utilisation}
\label{subsubsec:utilisation}

When configuring functions, developers typically specify the required memory, with other resources such as CPU shares and network bandwidth being allocated proportionally. However, due to the vast search space and performance variability, function configuration often becomes an ad-hoc decision-making process. This can lead to resource under-utilisation at both the function level and the underlying infrastructure, primarily as a result of resource over-provisioning \cite{Datadog2020}. Resource utilisation refers to the actual usage of allocated resources by a function during its execution lifecycle. For developers, this is critical because the interplay between allocated resources, function runtime cost, and performance is both complex and costly. From the perspective of CSPs, effective resource utilisation is equally important as it directly impacts their ability to optimise function placement and packing, ultimately reducing infrastructure wastage and improving operational efficiency.

Kim et al. \cite{CPUCap} propose a resource manager to control the CPU capacity for the functions to address the issue of resource contention. Their extensive set of experiments shows that their capacity control approach is able to reduce the resource contention by up to 44\% while not over-using the resources. On the other hand, Suresh et al. \cite{Ensure} regulate the resources for co-located workloads with a focus on reducing the costs of infrastructure at scale by as much as 52\%, compared to existing baselines. The analytical performance model by Mahmoudi and Khazaei \cite{PerformanceModellingSC} can be used for assessing different function configurations, helps understand various characteristics of the FaaS model and can benefit both the developer and CSP in resource planning. Moreover, Orion \cite{Orion} performs three optimisations on static DAGs, including function memory configuration, invocation bundling and function pre-warming to enhance resource utilisation while satisfying application SLA. In another work, Wang et al. \cite{EdgeAssistedVideoAnalytics} turn towards the configuration of serverless-based video analytics applications, where they investigate joint optimisation of use-case-specific controllable knobs and computation resource allocation to elevate resource utilisation. 

\subsubsection{Function Lifecycle: Time and Cost}
\label{subsubsec:latency}

FaaS functions ideally execute for a shorter duration, ranging from milliseconds to a few seconds, while the application expects a near real-time response. Existing research \cite{COSE.V2} consistently highlights that resource allocation significantly impacts the execution time of a function, eventually affecting the associated runtime costs. In particular, functions are found to speed up their execution with more resources, where serverless platforms usually tie function compute resources with memory allocation \cite{WithGreatFreedomComesGreatOpportunity}. This complicates the non-trivial configuration selection from a large search space where a complex interplay of runtime cost and execution time exists. Hence, a considerable amount of research efforts have been spent on optimising the function configuration process for refining execution time and runtime costs while maintaining the expected QoS and defined SLAs. However, the terms response time \cite{ModellingandOptimisationOfPerformanceCost}\cite{PredictingTheCostsOfWorkflows}, latency \cite{Costless}\cite{Llama}, job completion time \cite{Astrea} and execution time \cite{OFC}\cite{Lambdata} have been used interchangeably in the literature and we generalise them under \textit{function lifecycle time} and associated cost as \textit{function lifecycle cost}.

% \textcolor{red}{Should we define the terms Latency, Response time and Execution time?}

Function lifecycle time encapsulates the time taken across different stages of a function's execution, including the latency, execution time and end-to-end response time, and the lifetime cost reflects the corresponding cost based on the FaaS invocation model. To this end, Costless \cite{Costless} discusses the function lifecycle time and cost involved in serverless workflows and attempts to fuse a sequence of functions and configure appropriate memory for fused functions to reduce lifecycle cost with an acceptable increase in lifecycle time. Eismann et al. \cite{PredictingTheCostsOfWorkflows} identify that existing workflow cost estimation assumes a static function lifecycle time and emphasises on an input-sensitive analysis for predicting costs. The research \cite{COSE.V2} presents a statistical learning based framework that selects the near-optimal configurable parameter, function memory by utilising the predicted function lifecycle time and cost, both for single and workflow functions. From one point of view, Lin and Khazaei \cite{ModellingandOptimisationOfPerformanceCost} suggest that DAGs and PetriNets are effective in modelling serverless workflows for performance and propose a heuristic for predicting function lifecycle time and cost, if orchestration data is given. Alternatively, Mahmoudi and Khazaei  \cite{PerformanceModellingSC}\cite{PerformanceModellingMetricBased} propose analytical models that can be leveraged to enhance QoS and reduce costs by examining different serverless characteristics such as workload awareness, function configuration, and platform-specific metrics like container concurrency, etc. The performance models by \cite{Orion} estimate the end-to-end function lifecycle time for function resource allocation while reducing lifecycle time and cost.

Another study \cite{CostMinimisationSC} also found that function lifecycle time and cost are directly impacted by the configured resources and derived a profiling-based regression model for finding optimal memory settings and reducing costs. Zhu et al. \cite{RDOF} optimise the function deployment process and suggest a queuing network-based performance prediction model to find optimal configuration and reduce operating costs. Research works \cite{Llama}\cite{EdgeAssistedVideoAnalytics}\cite{charmseeker} target video analysis and processing pipelines to configure various resources like memory or CPU, including application-specific knobs such as the number of functions in a stage or video frames to reduce lifecycle time and associated operating costs. Sinha et al. \cite{Lachesis} focus on fine-grained per-request decoupled function configuration and leverage an online learning framework for a considerable lifecycle time speedup. Wen et al. \cite{StepConf} explore the degree of function parallelism in a serverless workflow for ensuring successful SLOs while reducing costs. While \cite{FireFace} propose to optimise the function configuration by exploiting the internal function feature with static code analysis, \cite{ParrotFish} proposes a function right-sizing framework based on online parametric regression to reduce the function operating costs while ensuring lifecycle time expectations. Similarly, the research \cite{UCCSiddharth} targets the function input size to make configuration decisions. The framework employs a regression model to predict the function configuration per request to reduce operational costs while maintaining running time constraints. Contrastingly, Wu et al. \cite{CEScaling} leverage serverless computing for ML model training and hyperparameter tuning, where they distribute the available resources among the different stages to reduce the overall completion time and costs. Unlike other works, \cite{SLOPE} aims at big data applications and utilises Graph Edit Distance (GDE)-based application similarity for resource estimation and workload prediction to improve operating costs. 

% \subsubsection{Execution cost}
% \label{subsubsec:cost}
\subsubsection{Data Locality and Data Exchange}
\label{subsubsec:locality}

A serverless function provides an abstraction for the users to deploy their business logic, where a CSP is responsible for resource allocation, function scheduling, networking and runtime management tasks. Recently, FaaS has been leveraged for a variety of application domains such as video processing \cite{Llama}\cite{charmseeker}, big data analytics \cite{Astrea}\cite{EvolutionaryProbabilistic}and ML training-inference \cite{Orion}. These applications generate large quantities of intermediate data, which is then processed in downstream stages. In the FaaS context, this requires a function-to-function exchange of this intermediate data, either wholly or partially. Existing research has identified that direct communication of functions is difficult \cite{Wisefuse} and usually leverage an external remote storage \cite{Lambdata} to promote the data or information exchange amongst functions. Therefore, several studies have identified data locality and exchange as a crucial FaaS challenge, and few have dedicated their efforts to understanding the impact of function configuration, along with its impact on function/application performance.

In study \cite{Lambdata}, researchers posit that there are data-aware caching and scheduling opportunities for serverless functions. To this end, the authors introduce a system to explicitly declare functions with data input and output locations to improve data locality and speed up function execution. Unlike this, the research conducted by Pandey and Kwon \cite{OptimizingMemoryAllocationinaServerless} discusses the over-allocation challenge faced in FaaS deployments and suggests that dependence on remote and external storage leads to exacerbated latency and reduced network bandwidth. It leverages this idea to consider image locality and execution history, and assigns similar functions to similar nodes to prioritise data-aware function execution. Research by Jarachanthan et al. \cite{Astrea} emphasise the difficulty involved in intermediate data exchange of data analytics applications and seeks optimally configured function parameters, such as degree of parallelism, resource allocation and number of stages, to deal with the complexity of ephemeral data management. Researchers in \cite{Llama} express that deployment frameworks should tune application-specific knobs such as sampling rate and batch size. They propose an automatic tuning framework that adjusts the degree of parallelism and resource configuration based on the intermediate output data generated from different stages.

\minrev{\textbf{Research Gaps:} Across the KPI dimensions surveyed, several gaps are evident. Resource utilisation is predominantly studied at the per-function level \minrev{\cite{CPUCap}\cite{PerformanceModellingSC}\cite{Ensure}}, with limited investigation into utilisation efficiency at the platform or multi-tenant infrastructure level under realistic bursty workloads. For function lifecycle cost and time, existing studies \minrev{\cite{Costless}\cite{PredictingTheCostsOfWorkflows}\cite{ModellingandOptimisationOfPerformanceCost}} rarely account for the joint effect of cold start latency and configuration changes simultaneously, particularly in the context of provisioned concurrency \minrev{\cite{Kraken}}. Data locality, while recognised as important, remains among the least addressed KPIs in configuration research; only a handful of works \cite{Lambdata}\cite{OptimizingMemoryAllocationinaServerless} explicitly optimise configuration for data-aware execution. Notably, energy consumption and carbon emissions \minrev{\cite{estimatingCarbonServerless}\cite{bridgingSustainabilityGap}\cite{hiddenCarbonServerless}\cite{accountableCarbonServerless}} have emerged as a new class of KPI that the current literature on function configuration has not yet systematically addressed, representing a significant gap explored in Section~\ref{subsubsec:sustainability}.}

\subsection{Resource Optimisation Targets}
\label{subsec:optimisedResource}

A serverless function's performance and runtime cost are directly influenced by its resource configuration. This configuration is typically managed in one of two ways: a coupled approach, where memory is the primary setting and other resources like CPU are proportionally allocated, typically in commercial offerings, or a decoupled approach, where resources can be configured independently, often seen in open-source platforms like OpenFaaS \cite{openfaas}. Finding the optimal resource configuration from a large search space is challenging, as different platforms expose multiple resource combinations that result in non-monotonic function performance. Therefore, a vast majority of existing works have focused on finding the right or optimal resource configuration to achieve a desired performance level. Therefore, existing research can be broadly categorised by the primary resource targeted for optimisation including memory, CPU, and function concurrency, in response to varying use-case requirements and performance goals.

\subsubsection{Coupled Resource Configuration}
\label{subsubsec:memory}

The current serverless operational model of AWS Lambda \cite{awsLambdaQuotas}, Azure Functions \cite{azureFunctions} and Cloud Run Functions \cite{gcfCPUAllocation} either offers memory as a single configurable knob or provides a fixed set of proportional resources to fine-tune function performance. For instance, in AWS Lambda, a function's memory setting decides the amount of CPU share, network bandwidth and I/O allocated to it during its execution. \rev{It is important to note, however, that this coupled model is evolving. Google Cloud Run (the container-serving product, distinct from Cloud Run \textit{Functions}) already supports independent CPU and memory allocation \cite{gcfCPUAllocation}, as does the broader Knative-based Serverless 2.0 ecosystem. Furthermore, in late 2025, AWS announced Lambda Managed Instances \cite{awsLambdaManagedInstances}, enabling Lambda functions to run on user-selected EC2 instance types with a configurable memory-to-vCPU ratio, effectively introducing decoupled resource control within the Lambda programming model. These developments indicate that the boundary between coupled and decoupled resource configuration is narrowing in production platforms.} Similarly, Cloud Run Functions allow different memory limits to be configured that correspond to a minimum amount of CPU. On the other hand, Azure Functions has various hosting plans that offer different virtual machine sizes where memory is metered dynamically. Therefore, memory plays a crucial role in the function lifecycle time and cost. One of the existing research \cite{Costless} targets memory configuration in a function workflow after fusing sequential functions at the cost of a 5\%-15\% increase in workflow latency. Akhtar et al. \cite{COSE.V2} find that a function's memory configuration directly impacts its runtime cost and execution delay. They propose a Bayesian Optimisation-based prediction method to select near-best configuration parameters to provide reduced runtime cost and satisfy user-defined delay constraints. In the study \cite{ModellingandOptimisationOfPerformanceCost}, analytical models are described to predict end-to-end execution time and cost on a critical path for serverless workflows. This work reduces DAGs into function chains and profiles the function performances at different memory configurations to select the one that satisfies either cost or time constraints. 

Mvondo et al. \cite{OFC} take a different approach towards function memory configuration and utilise the over-provisioned memory to fuel a cost-effective and fault-tolerant RAM-based caching system. Another research by Eismann et al. \cite{Sizeless} aims to predict near-optimal memory size of a function by utilising a Multi-Regression execution time model for all memory configurations. They leverage synthetic functions to develop execution time models and predict optimal memory for up to 79\% of functions while decreasing average runtime costs by 2.6\%. Zubko et al. \cite{MAFF} suggest search-based heuristics for memory optimisation to find the near-optimal function memory that satisfies the cost and runtime constraints. The researchers in \cite{charmseeker} exploit the relationship between the memory configuration, input workload and performance for serverless video processing pipelines. They suggest a Bayesian Optimisation-based stage-wise configuration finder that improves relative processing time by up to 408\% while satisfying workflow runtime budget. In another work by Jindal et al. \cite{SLAM}, memory configuration of workflow functions is demonstrated using a max heap data structure to meet user-specified SLO constraints. Their configuration finder iterated over different memory settings to generate execution time heap and utilise this information for SLO optimisation. 

According to the researchers \cite{Orion}, selecting the best VM size for a function in DAG is challenging as multiple resources are scaled orthogonal to memory, which is non-linear. Therefore, the study leverages per-function performance models that map VM sizes to latency distribution, utilise Best-First search to optimise for cost and select a memory configuration. The work presented in \cite{TimeCostMemoryConfig} builds on \cite{ModellingandOptimisationOfPerformanceCost} to suggest an urgency-based and meta-heuristic algorithm to optimise function workflow memory configuration under a specified budget. Cordingly et al. \cite{CPUTAMS} take a CPU time accounting approach, gather performance metrics, and utilise a regression model for predicting the desired function memory configuration. Pandey and Kwan \cite{OptimizingMemoryAllocationinaServerless} discuss resource wastage by memory over-provisioning and present a simulation-based study that estimates function memory to schedule similar functions on the same infrastructure nodes. The study \cite{EvolutionaryProbabilistic}, proposes an evolutionary genetic algorithm that utilises function memory configuration in a workflow to balance runtime performance and cost. Parrotfish \cite{ParrotFish} is a function right-sizing recommendation tool that takes advantage of online parametric regression for cost modelling and determines optimal function memory configuration within specified budget constraints. In another research, Agarwal et al. \cite{UCCSiddharth} take an input size-aware approach to configure functions and introduce Memfigless, which predicts performance and invokes functions with optimal memory to reduce running time costs.

\subsubsection{Decoupled Resource Configuration}
\label{subsubsec:cpu}

The CPU allocation in industrial serverless products is typically tied to memory configuration. As a result, a compute-intensive function or the one requiring additional CPU shares must be assigned more memory to obtain proportionally higher CPU resources. This often leads to resource over-provisioning and wastage while reducing overall utilisation. To address this, open-source serverless frameworks such as OpenFaaS, Knative and OpenWhisk provide decoupled resource allocation, allowing for flexible and refined function configuration. Moreover, numerous researches have criticised the constraints of coupled function resource allocation and have leveraged the decoupling to optimise function performance and runtime cost.

The authors \cite{CPUCap} emphasise on the resource contention observed by applications that have similar performance requirements. To this end, they develop a CPU capacity control mechanism to co-locate the workloads and minimise response time skewness. Similarly, Ensure \cite{Ensure} targets resource contention and posits that CPU availability for functions changes over time. Its resource manager dynamically regulates the CPU shares for functions executing on the same CPU core, categorising them based on runtime performance and resource demands. In \cite{Aquatope}, the authors highlight a performance variation between cold and warm functions that leads to different QoS and resource demands. To this end, a QoS and uncertainty-aware function-level decoupled resource manager is proposed for an end-to-end workflow that supports independent CPU-limit-based resource allocation. Bilal et al. \cite{WithGreatFreedomComesGreatOpportunity} also point out the inefficiencies of coupled resource allocation and harness black-box optimisation technique to evaluate the performance impact of decoupled CPU and memory configurations. Lachesis \cite{Lachesis} targets input characteristics and function semantics to suggest per-invocation function configuration. It specifically works for CPU allocation and employs supervised learning multi-class classification to predict the minimum number of required CPU cores per invocation to fulfil the SLOs.

\subsubsection{\rev{Sustainability-Oriented Configuration}}
\label{subsubsec:sustainability}

\rev{An emerging class of resource optimisation targets addresses energy consumption and carbon emissions as primary objectives for function configuration. While performance and cost have long been the dominant optimisation goals in FaaS, the environmental footprint of serverless computing is receiving growing attention from both researchers and practitioners. Function configuration choices including memory allocation, CPU provisioning, instance lifecycle settings, and concurrency parameters, directly determine the energy consumed per invocation, and by extension, the associated carbon emissions depending on the energy mix of the data centre region. Recent studies have begun to quantify this relationship. Roy et al.~\cite{hiddenCarbonServerless} identify significant hidden carbon costs in serverless platforms stemming from idle warm container retention and over-provisioning of resources, which are both directly controllable through configuration. Awwad et al.~\cite{estimatingCarbonServerless} present a methodology for estimating the carbon footprint of serverless functions on public cloud platforms, demonstrating that memory configuration and execution duration are the primary configuration levers affecting per-invocation emissions. Lin and Shahrad ~\cite{bridgingSustainabilityGap} propose a framework for carbon-aware pricing and observability in serverless environments, enabling developers to incorporate carbon cost as a first-class configuration objective. Sharma and Fuerst ~\cite{accountableCarbonServerless} further introduce energy profiling for serverless functions, linking function-level resource configuration directly to accountable energy and carbon metrics.}

\rev{These works collectively demonstrate that sustainability is not merely a platform-level concern but is intimately connected to per-function configuration decisions. As cloud providers increasingly commit to carbon-neutral operations and offer region-level carbon intensity signals, configuration strategies that jointly optimise for performance, cost, and carbon emissions represent a meaningful and timely research opportunity. This gap is further explored in Section~\ref{sec:futureres}.}

% \subsubsection{Function Concurrency}
% \label{subsubsec:concurrency}
\subsubsection{Other Specific Resources}
\label{subsubsec:otherResource}

FaaS find its relevance in a variety of application domains ranging from healthcare, finance, multimedia and IT. Serverless applications are becoming popular with use cases like IoT sensing, video streaming and processing, event-driven websites, AI/ML model training and inference, and Large Language Model (LLM) inference tasks. Typically, the performance and runtime cost of a serverless function vary with its allocated memory or proportional compute resources. However, use-case-specific configuration tuning and alternate configuration knobs have also been discussed in the existing literature, where distinct function characteristics are harnessed to enhance runtime performance or reduce operational cost.

Tang and Yang \cite{Lambdata} argue that current serverless platforms are oblivious to the data a function reads or writes that is immutable and miss out on data locality opportunities. To this end, a system of explicit declaration of function input/output object path is suggested to promote caching, data locality and speedup function execution. In \cite{Llama}, an automated video pipeline configuration framework, Llama, is demonstrated that tunes knobs like sampling rate and batch size in addition to function resource configuration per invocation. Similarly, the work \cite{EdgeAssistedVideoAnalytics} deals with video analytics specific factors such as frame rate and Deep Neural Network (DNN) selection in addition to computational resource configuration. Another work by Jindal et al. \cite{FnCapacity} estimates the function capacity, i.e., the maximum number of concurrent requests a function can handle before violating the SLA, at a set memory configuration and function concurrency settings. The authors argue that function capacities can help developers in both offline and online fashion to deploy functions with the right configurations. Wen et al. \cite{StepConf} analyse task parallelism as an additional configuration affecting performance for functions that can take advantage of multi-core CPU allocations. With this objective, inter (multi-threading) and intra (concurrent executions) function parallelism is considered for optimising function memory in a workflow. Along the same lines, the research \cite{Cypress} reorders and batches incoming requests to minimise resource consumption while maintaining a high degree of SLO based on the inputs received. Wu et al. \cite{CEScaling} target training and hyper-parameter tuning of serverless ML workflows to introduce cost-effective dynamic resource adjustment and partitioning across the different stages of ML workflow.

\subsection{Configuration and Management Strategy}
\label{subsec:strategy}

Serverless platforms employ various techniques and management strategies to simplify resource configuration for developers and users. A common approach seen in commercial offerings is a coupled resource model where developers configure a function's memory, and other resources like CPU are proportionally allocated but reduces the configuration flexibility. However, in contrast, some platforms offer a fixed set of resource tiers or decoupled configuration model where resources are configured independently. Moreover, the varying function timeouts and concurrency limits advertised across different platforms introduce additional complexity to resource allocation decisions. 

In the recent years, numerous scholarly works \cite{COSE.V1}\cite{Costless}\cite{SLAM}\cite{StepConf} as well as industry reports \cite{Datadog2020}\cite{Datadog} have highlighted the importance of optimal function configuration due to significant performance variation. These works suggest that, in general, function performance or execution speed increases with increased resources but eventually plateaus beyond a certain configuration due to the complex interplay of different resources. To this end, tools like AWS Lambda Power Tuning \cite{AWSPowerTuning} and AWS Compute Optimiser \cite{AWSComputeOpt} have been developed to analyse a function's past performance and recommend the most suitable function configuration. However, these solutions either profile the respective functions in isolation or need sufficient data points to recommend a preferred configuration, considering constraints like runtime cost and execution time. Accordingly, a number of configuration dimensions and approaches are explored in the research community to tackle and optimise the function configuration task. These can be broadly categorised under search-based and path-finding techniques, stochastic and approximate optimisation, ML-based approaches, and other specific modelling for function configuration tasks. 

\subsubsection{Search and Path-finding Techniques}
\label{subsubsec:searchBased}

In FaaS deployments, service providers usually require developers to specify minimum function resources for a successful execution. To achieve this, developers typically profile their functions at multiple configurations for computational efficiency and decide on the optimal one based on the desired time and cost constraints. However, navigating a vast configuration space and selecting the best setting is challenging due to inherent performance variations and profiling time. 

% \textit{Deterministic Search and Gradient-based Optimisation:} 
Accordingly, AWS Lambda Power Tuning tool \cite{AWSPowerTuning}, implemented as an AWS Step Functions workflow, runs the function across specified configurations (ranging from 128 MB to 10 GB) at a specified step-size for a minimum number of 5 iterations. This workflow measures the execution speed and cost at each set. Eventually, the tool analyses the execution logs, visualises the average results of each configuration in a chart, and recommends an optimal function configuration that balances the average execution time and runtime cost. In a comparable manner, Zubko et al. \cite{MAFF} propose a function memory allocation framework that analyses the execution logs either in active or passive mode to run the suggested function configuration. It uses \textit{linear search}, \textit{binary search}, and \textit{gradient descent} that search the memory configuration space to determine the best possible option and save it for future use. The proposed framework finds a trade-off between cost and performance and prioritises either the running costs or function performance for selecting optimal memory setting. 
% \textit{Path-finding Algorithms:} 
Furthermore, existing literature also explores deterministic graph-based search methods, such as path-finding algorithms, to address the resource optimisation in FaaS. Elgamal et al. \cite{Costless} propose to combine multiple sequential functions in a workflow and configure their memory configuration for successful execution. The study suggests that sometimes the cost of a workflow can be dominated by state transition cost, giving an opportunity to fuse functions and make a bigger one. To this end, the authors propose to generate a Cost Graph for the fused possibilities with various memory configurations and apply \textit{Dijkstra's Shortest Path} algorithm to find the best latency within the defined budget. Wen et al. \cite{StepConf} propose a workflow function configuration framework called StepConf. It leverages both inter- and intra-function parallelism to appropriately configure the workflow function memory allocation. By utilising \textit{Critical Path Algorithm}, StepConf determines the stage-wise latest completion time to optimise the function configurations. Taking an alternative approach, Jarachanthan et al. \cite{Astrea} address the resource configuration in data analytics or Map-Reduce style jobs. They develop function performance and cost models based on user requirements and formulate resource optimisation as a \textit{Shortest Path} problem in Graph theory. Tomaras et al. \cite{SLOPE} suggest that functions in a big-data application with similar characteristics, i.e., code-base, will have similar performance and resource requirements. To this end, they utilise \textit{Graph Edit Distance} to find similarity between the call graphs of two functions and utilise the prediction model of one, either wholly or partly, and provision a similar amount of resources.

% \textit{Heuristic \& Meta-Heuristic Techniques:} 
In addition to deterministic or exact search methods, heuristics-driven solutions employ certain rules or domain knowledge to guide their search process towards an optimal solution. To this end, Lin and Khazaei \cite{ModellingandOptimisationOfPerformanceCost} leverage performance and cost models to transform DAG-based function workflows into a linear structure and propose a heuristic called \textit{Probability Refined Critical Path} (PRCP) algorithm. The researchers utilise PRCP to optimise the function memory allocation for both budget-constrained performance and performance-constrained runtime cost by introducing different benefit-cost ratio strategies. Building on this work, Li et al. \cite{TimeCostMemoryConfig} simplify the DAG transformation and propose a heuristic \textit{Urgency-based Workflow Configuration} (UWC) to obtain a memory configuration under the budget constraints that minimises the execution time. Additionally, they employ \textit{Beetle Antennae Search} (BAS) to avoid locally optimal solutions and find a time-cost trade-off based memory allocation scheme. On the other hand, \cite{Orion} investigates request bundling, function right sizing, and pre-warming of static serverless DAGs. The research employs \textit{Best-First search} to optimise for function performance and cost based on user requirements, where it exploits interpolation to determine the performance and cost at various resource configurations.

\subsubsection{Stochastic and Approximate Optimisation}
\label{subsubsec:modellingBased}

A serverless environment, with its reduced resource management opportunities, has been a choice for bursty, embarrassingly parallel, and highly dynamic workloads. This dynamicity and unpredictability, along with performance and cost variation, present a unique resource configuration challenge. While deterministic search methods are simple and attempt to explore a large search space, they struggle to cope with the workload dynamicity, varying resource requirements, and cost-performance trade-offs. As a result, stochastic and approximation techniques have emerged as effective alternatives to fine-tune the function resource configuration by leveraging heuristics, meta-heuristics, probabilistic models, and adaptive learning mechanisms. These techniques enable efficient search and optimisation of function configuration while balancing fluctuating performance and cost.

% \textit{Bayesian Learning-based Optimisation: } 
In the work \cite{CEScaling}, researchers focus on resource allocation during hyperparameter tuning and model training in ML workflows. They utilise an Iterative Greedy resource allocation approach to partition the resources among different trial stages for hyperparameter tuning. Furthermore, Pareto boundary-based profiling estimation is leveraged to allocate resources that minimise cost and job completion time. Akhtar et al. \cite{COSE.V1} employ \textit{Bayesian Optimisation} (BO) to determine the black-box relationship between runtime cost or execution time and function memory configurations. The proposed framework learns a performance model by carefully sampling distinct memory configurations and leveraging Integer Linear Programming (ILP) to find the optimal configuration that minimises delay and cost constraints. Likewise, the researchers in \cite{WithGreatFreedomComesGreatOpportunity} explore surrogate model variations of BO and find Gaussian Process (GP) integrated BO superior in predicting the performance of untested resource configurations. Leveraging this information, the researchers suggest exposing resource configuration as either tunable knobs or an autonomous service by a CSP. In a separate work \cite{charmseeker}, the researchers posit that configuration tuning is exponentially non-trivial in a video processing pipeline due to a huge search space and high parallelism. This makes naive searching methods irrelevant and has higher search costs, which are not suitable for pipelines with many functions. To this end, \textit{Sequential BO} with a Gaussian surrogate model is applied to optimise both runtime cost and execution latency, which can get near-optimal function configuration with high probability. Researchers in \cite{Aquatope} assert that BO-influenced resource configuration can be negatively affected by the unpredictable interference inherent to FaaS environments. To mitigate this, the researchers introduce a noise-aware BO for function configuration at each workflow stage. It leverages a fixed noise GP surrogate model that incorporates QoS to predict function performance and filter candidate solutions.

% \textit{Adaptive Learning-based Optimisation: }
In addition to the Bayesian learning methods, \textit{evolutionary algorithms} (EA) are also explored for optimising the function configuration. These algorithms do not rely on a fixed path-finding approach; rather, they rely on randomised processes to search the configuration space with no intention to find the global optimum. Wang et al. \cite{EvolutionaryProbabilistic} criticise the poor infrastructure transparency of FaaS providers and aim to find the optimum balance between the runtime performance and cost via memory configuration. To this end, they propose an EA-based serverless workflow (EASW) configuration mechanism that leverages Polynomial Mutation for better offspring generation and optimises the memory for budget and performance-constrained objectives. Liu et al. \cite{FireFace} find that existing serverless platforms are progressively expanding towards the edge-cloud, where the limitation and heterogeneity of resources make their configuration challenging. The researchers propose FireFace, a configuration optimisation scheme that predicts execution time from internal function features such as input and output data size. Further, it utilises Adaptive Particle Swarm Optimisation with Genetic Algorithm (APSO-GA) to effectively search for a time and cost performant configuration setting. Supplementary to these works, Wang et al. \cite{EdgeAssistedVideoAnalytics} target the serverless surveillance video application and assert that there exists a non-linear relationship between the function configuration and KPIs. Further, they suggest a selection of object detection models and video knobs, such as input-based frame rate, for optimal video function configuration decisions. To this end, the authors formulate the cost-effective configuration problem as a stochastic process and find near-optimal solutions using \textit{Markov Approximation} with lower computational overhead.

\subsubsection{Machine Learning Techniques}
\label{subsubsec:learningBased}

FaaS has emerged as a transformative paradigm shift in how cloud resources are accessed and managed. As the serverless environments grow in popularity, adoption, and complexity, the traditional methods, such as search algorithms or optimisation techniques, may not hold up with the dynamism and unpredictability of workloads and function execution patterns. In comparison to heuristics or approximation methods, ML methods can learn function-specific execution patterns and apply a generalised model to unseen workloads, which enables efficient real-time configuration decisions. Therefore, ML-based techniques furnish an alternative approach to predict, optimise and dynamically configure function resources by utilising historical and runtime data. As a result, the integration of ML methods allows serverless platforms to advance towards enhanced system autonomy by intelligently and adaptively configuring function resources. This further reduces the developer effort and intervention in the configuration process while improving both runtime cost and performance. 

Eismann et al. \cite{PredictingTheCostsOfWorkflows} propose a learning model based on \textit{Mixture Density Network} (MDN) to predict response time and output parameter distribution, and utilise Monte-Carlo simulations for workflow cost prediction. By doing so, a developer could estimate the costs of running the workflow at various configurations and select the cheapest option. Spillner \cite{RMFunctionsProfiling} discusses the developer pain points where service providers still expose low-level function decisions like memory configuration. They highlight the need for tracing tools for measuring the dynamic function performance to enable informed memory configuration decisions and provide a learning-based auto-tuning tool for Docker-based function deployments. However, no further details of the model were provided. The researchers in \cite{OFC} target the complex relation of function performance and memory needs to power the in-memory cache for extract and load stages in an ETL pipeline. To this end, a \textit{J48 Decision Tree} implementation is proposed. The model extracts function features to predict and classify the upper bound on function memory requirement for successful execution at specific input parameters. Additionally, comparisons to Random Forest and Hoeffding Tree are also performed to find J48 model better in terms of recall, precision, and F-measure scores. 

A cost minimisation method is discussed by Sedefo\u{g}lu and S\"{o}zer in \cite{CostMinimisationSC} for deploying serverless functions. They posit that a function's memory has a significant impact on its runtime cost and derive a \textit{Regression Model} from profiling data at various memory configurations. This model is later utilised to search the configuration space for optimal memory settings. Similarly, Sizeless \cite{Sizeless} proposes to predict optimal function memory size by utilising synthetically generated invocation data and constructs an offline \textit{Multi-Target Regression} model based on profiling. The authors leverage this regression model in an online phase to estimate the function execution time at unseen configurations and select the optimal memory size. Agarwal et al. \cite{UCCSiddharth} present a \textit{Multi-output Random Forest Regression} model that considers function input size for predicting performance. The approach leverages Pareto front analysis for selecting, and executing the function with right or optimal memory configuration for the specific input while adhering to performance SLO and reducing operational cost. In another work \cite{CPUTAMS}, function CPU metrics are leveraged for regression modelling that maps vCPU utilised to function memory, predicting the memory configuration that offers the highest performance at the lowest cost. Pandey and Kwon in \cite{OptimizingMemoryAllocationinaServerless} discuss memory over-allocation and under-utilised resources that cause cold start and latency issues in function-based deployments. To address this, they propose a \textit{Random Forest}-based function memory estimator for functions with similar code dependencies, memory usage, etc., to schedule them on the same nodes and improve data locality. In the sudy \cite{ParrotFish}, researchers claim that right-sizing a function goes against the serverless philosophy and necessitates developer intervention for ad-hoc and experience-based function configuration. To this end, an Online Parametric Regression-based approach is outlined that fits the function performance data to a known family of mathematical functions and trains its model parameters. This helps in efficiently choosing the next exploration that maximises the information gain while minimising the cost.

Unlike regression modelling, Sinha et al. \cite{Lachesis} design resource configuration as a supervised learning problem. The \textit{Multi-Class Classification} model predicts the count of CPU cores required per function invocation to execute successfully with the lowest cost. A research by Jindal et al. \cite{FnCapacity} takes a different approach to approximate the number of function invocations an instance could serve before violating the SLO i.e., function capacity (FC). The study explores different regression models such as Linear, Polynomial, Ridge, and Random Forest (RF) regression, in addition to Deep Neural Network (DNN). In the experiments, DNN model captures the non-linear relationship of the parameters better than other regression techniques and therefore, is employed for estimating FC. 

\subsubsection{Formal and Analytical Modelling}
% Other Specific Modelling}
\label{subsubsec:optimisationBased}

While ML approaches, and stochastic and approximation techniques provide a flexible and adaptive way to configure function resources, there are other proposals that require explicit modelling based on application structure or workload characteristics. These methods do not fall under the previously defined categories and generally leverage domain knowledge and analytical or mathematical models to capture the relationship between function resources and their performance. 

Kim et al. \cite{CPUCap} address the challenge of resource contention where a large number of serverless functions are consolidated onto the multi-tenant CSP infrastructure. According to the researchers, this resource contention leads to a significant performance degradation with variable workload and displays a lack of responsiveness from CSP's resource manager. To this end, a dynamic \textit{CPU cap control} mechanism is introduced that adjusts the CPU usage limit of individual worker processes executing the functions based on Dask \cite{Dask} distributed computing implementation. The proposed solution categorises the applications into groups based on their performance requirements and observes the throttle time and queue length of workers to adjust the CPU limits. This dynamic adjustment ensures resource allocation fairness for executing functions while reducing contention and improving performance. Similarly, a sub-component of the resource manager proposed in \cite{Ensure} classifies the serverless applications as edge-triggered or massively parallel based on resource consumption and lifetime patterns to regulate the function CPU share. The study states that functions demonstrate different runtime characteristics and therefore require dynamic adjustment of CPU shares to tolerate resource contention. 

Unlike previous works, \cite{Lambdata} finds an opportunity to co-locate functions and improve data locality. The study suggests that function outputs are immutable, only adding new value to a new location owing to function idempotency. To this end, an explicit, pre-determined data intent declaration approach is proposed where the developers can specify the input and output location along with the function declaration. This allows easier function co-location, data caching and data locality optimisations while improving performance. Zhu et al. \cite{RDOF} describe a topology and orchestration specification for cloud applications (TOSCA) modelling approach to optimise the total operating cost while meeting performance goals. This model-driven approach employs Layered Queuing Network (LQN) for performance prediction at different configurations and utilises a Genetic algorithm for the cheapest deployment scheme. The study \cite{Llama} posits that users are still required to manually and exhaustively tune and configure the resources in a video analytics application. Further, it is identified that large configuration space, input-dependent workflow execution and dynamic adjustment of per invocation resources are the major challenges in the video auto-tuning process. For this purpose, a collaboration of techniques such as dynamic slack calculation, safe delayed batching, early speculation, late commit and priority-based commit is proposed that automatically tunes each invocation for pipeline latency targets while minimising cost. One of the research by Mahmoudi and Khazaei \cite{PerformanceModellingSC} presents an analytical performance model, Semi-Markov Process, to help the CSPs and users to understand the different characteristics of the serverless platform in a steady state that could help them reduce the resource costs and improve QoS. They formulate the model based on \(M/M/m/m\) queuing theory and infer that different function expiration settings enable CSPs and users to assess various configurations for their performance and runtime cost. They further state that straightforward profiling results at different configurations can be used to form these models and later can be compared for a cheaper or more performant setting. Another work by them \cite{PerformanceModellingMetricBased} attempts to provide accurate analytical models to predict the function concurrency value, inspired by Knative \cite{knative}, and rate of requests per second (RPS) for autoscaling decisions. The authors formulate the metric estimation models to predict the performance and cost of deployments with different configurations that enable CSPs and developers to fine-tune their function configuration and select the best outcome.

\minrev{\textbf{Research Gaps:} Across configuration strategy techniques, several gaps merit future investigation. Search and path-finding methods are predominantly evaluated on static DAGs \minrev{\cite{TimeCostMemoryConfig}\cite{ModellingandOptimisationOfPerformanceCost}} and benchmark workloads while their scalability and applicability to dynamic, highly concurrent workloads with varying concurrency limits remain underexplored. Stochastic and Bayesian optimisation approaches \minrev{\cite{COSE.V1}\cite{charmseeker}\cite{Aquatope}} incur exploration costs that can be prohibitive for short-lived or infrequently invoked functions, thus, lightweight surrogate strategies for these cases are lacking. Among ML techniques \minrev{\cite{UCCSiddharth}\cite{Sizeless}\cite{ParrotFish}}, RL has received limited attention in the FaaS configuration context  relative to its potential for online adaptive tuning. Formal and analytical models \minrev{\cite{PerformanceModellingSC}\cite{PerformanceModellingMetricBased}\cite{RDOF}} largely assume steady-state conditions and are rarely validated against production-scale traces. Finally, no existing configuration strategy in the surveyed literature jointly optimises for performance, cost, and energy or carbon efficiency \minrev{\cite{estimatingCarbonServerless}\cite{bridgingSustainabilityGap}\cite{hiddenCarbonServerless}\cite{accountableCarbonServerless}}, representing an important direction for future work as discussed in Section~\ref{sec:futureres}.}

\section{Classification of Function Configuration Techniques using Taxonomy}
\label{sec:classification}

In this section, we review existing key works on function configuration that identify most
with the proposed taxonomy. In essence, we present here the works that explore novel techniques
for one or more of the primary aspects of function configuration that we have identified.

\begin{landscape}
% \tiny
\fontsize{6.5}{7.5}\selectfont
\setlength{\tabcolsep}{2.5pt}
\renewcommand{\arraystretch}{0.95}
\begin{longtable}{|p{0.6cm}|c|cc|cc|ccc|cccc|cccc|}
\caption{\minrev{Classification of surveyed function configuration techniques by taxonomy dimensions}}
\label{tab:related-works}\\
\hline
\multirow{2}{*}{Work} & \multirow{2}{*}{Year} & \multicolumn{2}{c|}{\begin{tabular}[c]{@{}c@{}}Workload\\ Characterisation\end{tabular}} & \multicolumn{2}{c|}{\begin{tabular}[c]{@{}c@{}}Deployment\\ Environment\end{tabular}} & \multicolumn{3}{c|}{KPI} & \multicolumn{4}{c|}{\begin{tabular}[c]{@{}c@{}}Resource Optimisation\\ Targets\end{tabular}} & \multicolumn{4}{c|}{Strategy} \\ \cline{3-17}
 &  & \multicolumn{1}{c|}{\begin{tabular}[c]{@{}c@{}}Single \\ Function\end{tabular}} & \begin{tabular}[c]{@{}c@{}c@{}}Workflow/\\ Function \\ Chain\end{tabular} & \multicolumn{1}{c|}{Commercial} & \multicolumn{1}{c|}{\begin{tabular}[c]{@{}c@{}}Open \\ Source\end{tabular}} & \multicolumn{1}{c|}{\begin{tabular}[c]{@{}c@{}}Resource \\ Utilisation\end{tabular}} & \multicolumn{1}{c|}{\begin{tabular}[c]{@{}c@{}}Function \\ Lifecycle\end{tabular}} & \begin{tabular}[c]{@{}c@{}c@{}}Data \\ Locality \&\\ Exchange\end{tabular} & \multicolumn{1}{c|}{\begin{tabular}[c]{@{}c@{}}Coupled\\ (Memory)\end{tabular}} & \multicolumn{1}{c|}{Decoupled} & \multicolumn{1}{c|}{\begin{tabular}[c]{@{}c@{}}Sustainability\\ Oriented\end{tabular}} & \begin{tabular}[c]{@{}c@{}}Application \\ Specific\end{tabular} & \multicolumn{1}{c|}{\begin{tabular}[c]{@{}c@{}c@{}}Search \&\\ Path \\Finding\end{tabular}} & \multicolumn{1}{c|}{\begin{tabular}[c]{@{}c@{}}Stochastic \&\\ Approximate \\ Optimisation\end{tabular}} & \multicolumn{1}{c|}{ML} & \begin{tabular}[c]{@{}c@{}}Formal \&\\ Analytical\\ Modelling\end{tabular} \\ \hline
\endhead
%
% ROW 1 — grey
\rowcolor{rowgray}
\cite{UCCSiddharth} & 2024 & \multicolumn{1}{c|}{\ding{51}} & \ding{55} & \multicolumn{1}{c|}{\ding{51}} & \ding{55} & \multicolumn{1}{c|}{\ding{51}} & \multicolumn{1}{c|}{\ding{51}} & \ding{55} & \multicolumn{1}{c|}{\ding{55}} & \multicolumn{1}{c|}{\ding{55}} & \multicolumn{1}{c|}{\ding{55}} & \ding{55} & \multicolumn{1}{c|}{\ding{55}} & \multicolumn{1}{c|}{\ding{55}} & \multicolumn{1}{c|}{\ding{51}} & \ding{55} \\ \hline
% ROW 2 — white
\cite{COSE.V1} & 2020 & \multicolumn{1}{c|}{\ding{51}} & \ding{55} & \multicolumn{1}{c|}{\ding{55}} & \ding{55} & \multicolumn{1}{c|}{\ding{55}} & \multicolumn{1}{c|}{\ding{55}} & \ding{55} & \multicolumn{1}{c|}{\ding{51}} & \multicolumn{1}{c|}{\ding{55}} & \multicolumn{1}{c|}{\ding{55}} & \ding{55} & \multicolumn{1}{c|}{\ding{55}} & \multicolumn{1}{c|}{\ding{51}} & \multicolumn{1}{c|}{\ding{55}} & \ding{55} \\ \hline
% ROW 3 — grey
\rowcolor{rowgray}
\cite{AWSComputeOpt} & 2024 & \multicolumn{1}{c|}{\ding{51}} & \ding{55} & \multicolumn{1}{c|}{\ding{51}} & \ding{55} & \multicolumn{1}{c|}{\ding{55}} & \multicolumn{1}{c|}{\ding{55}} & \ding{55} & \multicolumn{1}{c|}{\ding{55}} & \multicolumn{1}{c|}{\ding{55}} & \multicolumn{1}{c|}{\ding{55}} & \ding{55} & \multicolumn{1}{c|}{\ding{55}} & \multicolumn{1}{c|}{\ding{55}} & \multicolumn{1}{c|}{\ding{55}} & \ding{55} \\ \hline
% ROW 4 — white
\cite{Cypress} & 2022 & \multicolumn{1}{c|}{\ding{51}} & \ding{55} & \multicolumn{1}{c|}{\ding{55}} & \ding{55} & \multicolumn{1}{c|}{\ding{55}} & \multicolumn{1}{c|}{\ding{55}} & \ding{55} & \multicolumn{1}{c|}{\ding{55}} & \multicolumn{1}{c|}{\ding{55}} & \multicolumn{1}{c|}{\ding{55}} & \ding{51} & \multicolumn{1}{c|}{\ding{55}} & \multicolumn{1}{c|}{\ding{55}} & \multicolumn{1}{c|}{\ding{55}} & \ding{55} \\ \hline
% ROW 5 — grey
\rowcolor{rowgray}
\cite{Kraken} & 2021 & \multicolumn{1}{c|}{\ding{55}} & \ding{51} & \multicolumn{1}{c|}{\ding{55}} & \ding{55} & \multicolumn{1}{c|}{\ding{55}} & \multicolumn{1}{c|}{\ding{55}} & \ding{55} & \multicolumn{1}{c|}{\ding{55}} & \multicolumn{1}{c|}{\ding{55}} & \multicolumn{1}{c|}{\ding{55}} & \ding{55} & \multicolumn{1}{c|}{\ding{55}} & \multicolumn{1}{c|}{\ding{55}} & \multicolumn{1}{c|}{\ding{55}} & \ding{55} \\ \hline
% ROW 6 — white
\cite{WithGreatFreedomComesGreatOpportunity} & 2023 & \multicolumn{1}{c|}{\ding{55}} & \ding{51} & \multicolumn{1}{c|}{\ding{55}} & \ding{55} & \multicolumn{1}{c|}{\ding{55}} & \multicolumn{1}{c|}{\ding{51}} & \ding{55} & \multicolumn{1}{c|}{\ding{55}} & \multicolumn{1}{c|}{\ding{51}} & \multicolumn{1}{c|}{\ding{55}} & \ding{55} & \multicolumn{1}{c|}{\ding{55}} & \multicolumn{1}{c|}{\ding{51}} & \multicolumn{1}{c|}{\ding{55}} & \ding{55} \\ \hline
% ROW 7 — grey
\rowcolor{rowgray}
\cite{AWSPowerTuning} & 2023 & \multicolumn{1}{c|}{\ding{51}} & \ding{55} & \multicolumn{1}{c|}{\ding{51}} & \ding{55} & \multicolumn{1}{c|}{\ding{55}} & \multicolumn{1}{c|}{\ding{55}} & \ding{55} & \multicolumn{1}{c|}{\ding{55}} & \multicolumn{1}{c|}{\ding{55}} & \multicolumn{1}{c|}{\ding{55}} & \ding{55} & \multicolumn{1}{c|}{\ding{51}} & \multicolumn{1}{c|}{\ding{55}} & \multicolumn{1}{c|}{\ding{55}} & \ding{55} \\ \hline
% ROW 8 — white
\cite{CPUTAMS} & 2022 & \multicolumn{1}{c|}{\ding{51}} & \ding{55} & \multicolumn{1}{c|}{\ding{51}} & \ding{51} & \multicolumn{1}{c|}{\ding{55}} & \multicolumn{1}{c|}{\ding{55}} & \ding{55} & \multicolumn{1}{c|}{\ding{51}} & \multicolumn{1}{c|}{\ding{55}} & \multicolumn{1}{c|}{\ding{55}} & \ding{55} & \multicolumn{1}{c|}{\ding{55}} & \multicolumn{1}{c|}{\ding{55}} & \multicolumn{1}{c|}{\ding{51}} & \ding{55} \\ \hline
% ROW 9 — grey
\rowcolor{rowgray}
\cite{Sizeless} & 2021 & \multicolumn{1}{c|}{\ding{51}} & \ding{55} & \multicolumn{1}{c|}{\ding{51}} & \ding{55} & \multicolumn{1}{c|}{\ding{51}} & \multicolumn{1}{c|}{\ding{55}} & \ding{55} & \multicolumn{1}{c|}{\ding{55}} & \multicolumn{1}{c|}{\ding{55}} & \multicolumn{1}{c|}{\ding{55}} & \ding{55} & \multicolumn{1}{c|}{\ding{55}} & \multicolumn{1}{c|}{\ding{55}} & \multicolumn{1}{c|}{\ding{51}} & \ding{55} \\ \hline
% ROW 10 — white
\cite{PredictingTheCostsOfWorkflows} & 2020 & \multicolumn{1}{c|}{\ding{55}} & \ding{51} & \multicolumn{1}{c|}{\ding{51}} & \ding{55} & \multicolumn{1}{c|}{\ding{55}} & \multicolumn{1}{c|}{\ding{51}} & \ding{55} & \multicolumn{1}{c|}{\ding{55}} & \multicolumn{1}{c|}{\ding{55}} & \multicolumn{1}{c|}{\ding{55}} & \ding{55} & \multicolumn{1}{c|}{\ding{55}} & \multicolumn{1}{c|}{\ding{55}} & \multicolumn{1}{c|}{\ding{51}} & \ding{55} \\ \hline
% ROW 11 — grey
\rowcolor{rowgray}
\cite{Costless} & 2018 & \multicolumn{1}{c|}{\ding{55}} & \ding{51} & \multicolumn{1}{c|}{\ding{51}} & \ding{51} & \multicolumn{1}{c|}{\ding{55}} & \multicolumn{1}{c|}{\ding{51}} & \ding{55} & \multicolumn{1}{c|}{\ding{51}} & \multicolumn{1}{c|}{\ding{55}} & \multicolumn{1}{c|}{\ding{55}} & \ding{55} & \multicolumn{1}{c|}{\ding{51}} & \multicolumn{1}{c|}{\ding{55}} & \multicolumn{1}{c|}{\ding{55}} & \ding{55} \\ \hline
% ROW 12 — white
\cite{Astrea} & 2022 & \multicolumn{1}{c|}{\ding{55}} & \ding{51} & \multicolumn{1}{c|}{\ding{51}} & \ding{55} & \multicolumn{1}{c|}{\ding{55}} & \multicolumn{1}{c|}{\ding{51}} & \ding{51} & \multicolumn{1}{c|}{\ding{55}} & \multicolumn{1}{c|}{\ding{55}} & \multicolumn{1}{c|}{\ding{55}} & \ding{55} & \multicolumn{1}{c|}{\ding{51}} & \multicolumn{1}{c|}{\ding{55}} & \multicolumn{1}{c|}{\ding{55}} & \ding{55} \\ \hline
% ROW 13 — grey
\rowcolor{rowgray}
\cite{FnCapacity} & 2022 & \multicolumn{1}{c|}{\ding{55}} & \ding{55} & \multicolumn{1}{c|}{\ding{55}} & \ding{55} & \multicolumn{1}{c|}{\ding{55}} & \multicolumn{1}{c|}{\ding{55}} & \ding{55} & \multicolumn{1}{c|}{\ding{55}} & \multicolumn{1}{c|}{\ding{55}} & \multicolumn{1}{c|}{\ding{55}} & \ding{51} & \multicolumn{1}{c|}{\ding{55}} & \multicolumn{1}{c|}{\ding{55}} & \multicolumn{1}{c|}{\ding{51}} & \ding{55} \\ \hline
% ROW 14 — white
\cite{CPUCap} & 2020 & \multicolumn{1}{c|}{\ding{51}} & \ding{55} & \multicolumn{1}{c|}{\ding{55}} & \ding{55} & \multicolumn{1}{c|}{\ding{51}} & \multicolumn{1}{c|}{\ding{55}} & \ding{55} & \multicolumn{1}{c|}{\ding{55}} & \multicolumn{1}{c|}{\ding{51}} & \multicolumn{1}{c|}{\ding{55}} & \ding{55} & \multicolumn{1}{c|}{\ding{55}} & \multicolumn{1}{c|}{\ding{55}} & \multicolumn{1}{c|}{\ding{55}} & \ding{51} \\ \hline
% ROW 15 — grey
\rowcolor{rowgray}
\cite{FireFace} & 2023 & \multicolumn{1}{c|}{\ding{51}} & \ding{55} & \multicolumn{1}{c|}{\ding{55}} & \ding{51} & \multicolumn{1}{c|}{\ding{55}} & \multicolumn{1}{c|}{\ding{51}} & \ding{55} & \multicolumn{1}{c|}{\ding{55}} & \multicolumn{1}{c|}{\ding{55}} & \multicolumn{1}{c|}{\ding{55}} & \ding{55} & \multicolumn{1}{c|}{\ding{55}} & \multicolumn{1}{c|}{\ding{51}} & \multicolumn{1}{c|}{\ding{55}} & \ding{55} \\ \hline
% ROW 16 — white
\cite{TimeCostMemoryConfig} & 2022 & \multicolumn{1}{c|}{\ding{55}} & \ding{55} & \multicolumn{1}{c|}{\ding{55}} & \ding{55} & \multicolumn{1}{c|}{\ding{55}} & \multicolumn{1}{c|}{\ding{55}} & \ding{55} & \multicolumn{1}{c|}{\ding{51}} & \multicolumn{1}{c|}{\ding{55}} & \multicolumn{1}{c|}{\ding{55}} & \ding{55} & \multicolumn{1}{c|}{\ding{51}} & \multicolumn{1}{c|}{\ding{55}} & \multicolumn{1}{c|}{\ding{55}} & \ding{55} \\ \hline
% ROW 17 — grey
\rowcolor{rowgray}
\cite{ModellingandOptimisationOfPerformanceCost} & 2020 & \multicolumn{1}{c|}{\ding{55}} & \ding{51} & \multicolumn{1}{c|}{\ding{51}} & \ding{55} & \multicolumn{1}{c|}{\ding{55}} & \multicolumn{1}{c|}{\ding{51}} & \ding{55} & \multicolumn{1}{c|}{\ding{51}} & \multicolumn{1}{c|}{\ding{55}} & \multicolumn{1}{c|}{\ding{55}} & \ding{55} & \multicolumn{1}{c|}{\ding{51}} & \multicolumn{1}{c|}{\ding{55}} & \multicolumn{1}{c|}{\ding{55}} & \ding{55} \\ \hline
% ROW 18 — white
\cite{Orion} & 2022 & \multicolumn{1}{c|}{\ding{55}} & \ding{51} & \multicolumn{1}{c|}{\ding{55}} & \ding{55} & \multicolumn{1}{c|}{\ding{51}} & \multicolumn{1}{c|}{\ding{51}} & \ding{55} & \multicolumn{1}{c|}{\ding{51}} & \multicolumn{1}{c|}{\ding{55}} & \multicolumn{1}{c|}{\ding{55}} & \ding{55} & \multicolumn{1}{c|}{\ding{51}} & \multicolumn{1}{c|}{\ding{55}} & \multicolumn{1}{c|}{\ding{55}} & \ding{55} \\ \hline
% ROW 19 — grey
\rowcolor{rowgray}
\cite{Wisefuse} & 2022 & \multicolumn{1}{c|}{\ding{55}} & \ding{55} & \multicolumn{1}{c|}{\ding{55}} & \ding{55} & \multicolumn{1}{c|}{\ding{55}} & \multicolumn{1}{c|}{\ding{55}} & \ding{55} & \multicolumn{1}{c|}{\ding{55}} & \multicolumn{1}{c|}{\ding{55}} & \multicolumn{1}{c|}{\ding{55}} & \ding{55} & \multicolumn{1}{c|}{\ding{55}} & \multicolumn{1}{c|}{\ding{55}} & \multicolumn{1}{c|}{\ding{55}} & \ding{55} \\ \hline
% ROW 20 — white
\cite{PerformanceModellingSC} & 2022 & \multicolumn{1}{c|}{\ding{51}} & \ding{51} & \multicolumn{1}{c|}{\ding{51}} & \ding{55} & \multicolumn{1}{c|}{\ding{51}} & \multicolumn{1}{c|}{\ding{51}} & \ding{55} & \multicolumn{1}{c|}{\ding{55}} & \multicolumn{1}{c|}{\ding{55}} & \multicolumn{1}{c|}{\ding{55}} & \ding{55} & \multicolumn{1}{c|}{\ding{55}} & \multicolumn{1}{c|}{\ding{55}} & \multicolumn{1}{c|}{\ding{55}} & \ding{51} \\ \hline
% ROW 21 — grey
\rowcolor{rowgray}
\cite{PerformanceModellingMetricBased} & 2023 & \multicolumn{1}{c|}{\ding{55}} & \ding{55} & \multicolumn{1}{c|}{\ding{55}} & \ding{55} & \multicolumn{1}{c|}{\ding{55}} & \multicolumn{1}{c|}{\ding{51}} & \ding{55} & \multicolumn{1}{c|}{\ding{55}} & \multicolumn{1}{c|}{\ding{55}} & \multicolumn{1}{c|}{\ding{55}} & \ding{55} & \multicolumn{1}{c|}{\ding{55}} & \multicolumn{1}{c|}{\ding{55}} & \multicolumn{1}{c|}{\ding{55}} & \ding{51} \\ \hline
% ROW 22 — white
\cite{ParrotFish} & 2023 & \multicolumn{1}{c|}{\ding{51}} & \ding{55}\rev{$^\dagger$} & \multicolumn{1}{c|}{\ding{51}} & \ding{55} & \multicolumn{1}{c|}{\ding{55}} & \multicolumn{1}{c|}{\ding{51}} & \ding{55} & \multicolumn{1}{c|}{\ding{51}} & \multicolumn{1}{c|}{\ding{55}} & \multicolumn{1}{c|}{\ding{55}} & \ding{55} & \multicolumn{1}{c|}{\ding{55}} & \multicolumn{1}{c|}{\ding{55}} & \multicolumn{1}{c|}{\ding{51}} & \ding{55} \\ \hline
% ROW 23 — grey
\rowcolor{rowgray}
\cite{OFC} & 2021 & \multicolumn{1}{c|}{\ding{51}} & \ding{55} & \multicolumn{1}{c|}{\ding{55}} & \ding{51} & \multicolumn{1}{c|}{\ding{55}} & \multicolumn{1}{c|}{\ding{51}} & \ding{55} & \multicolumn{1}{c|}{\ding{51}} & \multicolumn{1}{c|}{\ding{55}} & \multicolumn{1}{c|}{\ding{55}} & \ding{55} & \multicolumn{1}{c|}{\ding{55}} & \multicolumn{1}{c|}{\ding{55}} & \multicolumn{1}{c|}{\ding{51}} & \ding{55} \\ \hline
% ROW 24 — white
\cite{OptimizingMemoryAllocationinaServerless} & 2023 & \multicolumn{1}{c|}{\ding{51}} & \ding{51} & \multicolumn{1}{c|}{\ding{55}} & \ding{55} & \multicolumn{1}{c|}{\ding{55}} & \multicolumn{1}{c|}{\ding{55}} & \ding{51} & \multicolumn{1}{c|}{\ding{51}} & \multicolumn{1}{c|}{\ding{55}} & \multicolumn{1}{c|}{\ding{55}} & \ding{55} & \multicolumn{1}{c|}{\ding{55}} & \multicolumn{1}{c|}{\ding{55}} & \multicolumn{1}{c|}{\ding{51}} & \ding{55} \\ \hline
% ROW 25 — grey
\rowcolor{rowgray}
\cite{COSE.V2} & 2023 & \multicolumn{1}{c|}{\ding{51}} & \ding{55} & \multicolumn{1}{c|}{\ding{51}} & \ding{55} & \multicolumn{1}{c|}{\ding{55}} & \multicolumn{1}{c|}{\ding{51}} & \ding{55} & \multicolumn{1}{c|}{\ding{51}} & \multicolumn{1}{c|}{\ding{55}} & \multicolumn{1}{c|}{\ding{55}} & \ding{55} & \multicolumn{1}{c|}{\ding{55}} & \multicolumn{1}{c|}{\ding{51}} & \multicolumn{1}{c|}{\ding{55}} & \ding{55} \\ \hline
% ROW 26 — white
\cite{Llama} & 2021 & \multicolumn{1}{c|}{\ding{55}} & \ding{51} & \multicolumn{1}{c|}{\ding{51}} & \ding{55} & \multicolumn{1}{c|}{\ding{55}} & \multicolumn{1}{c|}{\ding{51}} & \ding{51} & \multicolumn{1}{c|}{\ding{51}} & \multicolumn{1}{c|}{\ding{55}} & \multicolumn{1}{c|}{\ding{55}} & \ding{51} & \multicolumn{1}{c|}{\ding{55}} & \multicolumn{1}{c|}{\ding{55}} & \multicolumn{1}{c|}{\ding{55}} & \ding{51} \\ \hline
% ROW 27 — grey
\rowcolor{rowgray}
\cite{SLAM} & 2022 & \multicolumn{1}{c|}{\ding{55}} & \ding{51} & \multicolumn{1}{c|}{\ding{55}} & \ding{55} & \multicolumn{1}{c|}{\ding{55}} & \multicolumn{1}{c|}{\ding{55}} & \ding{55} & \multicolumn{1}{c|}{\ding{51}} & \multicolumn{1}{c|}{\ding{55}} & \multicolumn{1}{c|}{\ding{55}} & \ding{55} & \multicolumn{1}{c|}{\ding{55}} & \multicolumn{1}{c|}{\ding{55}} & \multicolumn{1}{c|}{\ding{55}} & \ding{55} \\ \hline
% ROW 28 — white
\cite{CostMinimisationSC} & 2021 & \multicolumn{1}{c|}{\ding{55}} & \ding{55} & \multicolumn{1}{c|}{\ding{55}} & \ding{55} & \multicolumn{1}{c|}{\ding{55}} & \multicolumn{1}{c|}{\ding{51}} & \ding{55} & \multicolumn{1}{c|}{\ding{55}} & \multicolumn{1}{c|}{\ding{55}} & \multicolumn{1}{c|}{\ding{55}} & \ding{55} & \multicolumn{1}{c|}{\ding{55}} & \multicolumn{1}{c|}{\ding{55}} & \multicolumn{1}{c|}{\ding{55}} & \ding{55} \\ \hline
% ROW 29 — grey
\rowcolor{rowgray}
\cite{Latest} & 2022 & \multicolumn{1}{c|}{\ding{55}} & \ding{55} & \multicolumn{1}{c|}{\ding{55}} & \ding{51} & \multicolumn{1}{c|}{\ding{55}} & \multicolumn{1}{c|}{\ding{55}} & \ding{55} & \multicolumn{1}{c|}{\ding{55}} & \multicolumn{1}{c|}{\ding{55}} & \multicolumn{1}{c|}{\ding{55}} & \ding{55} & \multicolumn{1}{c|}{\ding{55}} & \multicolumn{1}{c|}{\ding{55}} & \multicolumn{1}{c|}{\ding{55}} & \ding{55} \\ \hline
% ROW 30 — white
\cite{Lachesis} & 2023 & \multicolumn{1}{c|}{\ding{55}} & \ding{55} & \multicolumn{1}{c|}{\ding{55}} & \ding{51} & \multicolumn{1}{c|}{\ding{55}} & \multicolumn{1}{c|}{\ding{51}} & \ding{55} & \multicolumn{1}{c|}{\ding{55}} & \multicolumn{1}{c|}{\ding{51}} & \multicolumn{1}{c|}{\ding{55}} & \ding{55} & \multicolumn{1}{c|}{\ding{55}} & \multicolumn{1}{c|}{\ding{55}} & \multicolumn{1}{c|}{\ding{51}} & \ding{55} \\ \hline
% ROW 31 — grey
\rowcolor{rowgray}
\cite{RMFunctionsProfiling} & 2020 & \multicolumn{1}{c|}{\ding{51}} & \ding{55} & \multicolumn{1}{c|}{\ding{55}} & \ding{55} & \multicolumn{1}{c|}{\ding{55}} & \multicolumn{1}{c|}{\ding{55}} & \ding{55} & \multicolumn{1}{c|}{\ding{55}} & \multicolumn{1}{c|}{\ding{55}} & \multicolumn{1}{c|}{\ding{55}} & \ding{55} & \multicolumn{1}{c|}{\ding{55}} & \multicolumn{1}{c|}{\ding{55}} & \multicolumn{1}{c|}{\ding{51}} & \ding{55} \\ \hline
% ROW 32 — white
\cite{Ensure} & 2020 & \multicolumn{1}{c|}{\ding{51}} & \ding{55} & \multicolumn{1}{c|}{\ding{55}} & \ding{55} & \multicolumn{1}{c|}{\ding{51}} & \multicolumn{1}{c|}{\ding{55}} & \ding{55} & \multicolumn{1}{c|}{\ding{55}} & \multicolumn{1}{c|}{\ding{51}} & \multicolumn{1}{c|}{\ding{55}} & \ding{55} & \multicolumn{1}{c|}{\ding{55}} & \multicolumn{1}{c|}{\ding{55}} & \multicolumn{1}{c|}{\ding{55}} & \ding{51} \\ \hline
% ROW 33 — grey
\rowcolor{rowgray}
\cite{Lambdata} & 2020 & \multicolumn{1}{c|}{\ding{51}} & \ding{55} & \multicolumn{1}{c|}{\ding{55}} & \ding{51} & \multicolumn{1}{c|}{\ding{55}} & \multicolumn{1}{c|}{\ding{51}} & \ding{51} & \multicolumn{1}{c|}{\ding{55}} & \multicolumn{1}{c|}{\ding{55}} & \multicolumn{1}{c|}{\ding{55}} & \ding{51} & \multicolumn{1}{c|}{\ding{55}} & \multicolumn{1}{c|}{\ding{55}} & \multicolumn{1}{c|}{\ding{55}} & \ding{51} \\ \hline
% ROW 34 — white
\cite{SLOPE} & 2023 & \multicolumn{1}{c|}{\ding{55}} & \ding{55} & \multicolumn{1}{c|}{\ding{55}} & \ding{55} & \multicolumn{1}{c|}{\ding{55}} & \multicolumn{1}{c|}{\ding{51}} & \ding{55} & \multicolumn{1}{c|}{\ding{55}} & \multicolumn{1}{c|}{\ding{55}} & \multicolumn{1}{c|}{\ding{55}} & \ding{55} & \multicolumn{1}{c|}{\ding{51}} & \multicolumn{1}{c|}{\ding{55}} & \multicolumn{1}{c|}{\ding{55}} & \ding{55} \\ \hline
% ROW 35 — grey
\rowcolor{rowgray}
\cite{EvolutionaryProbabilistic} & 2023 & \multicolumn{1}{c|}{\ding{55}} & \ding{51} & \multicolumn{1}{c|}{\ding{55}} & \ding{55} & \multicolumn{1}{c|}{\ding{55}} & \multicolumn{1}{c|}{\ding{55}} & \ding{55} & \multicolumn{1}{c|}{\ding{51}} & \multicolumn{1}{c|}{\ding{55}} & \multicolumn{1}{c|}{\ding{55}} & \ding{55} & \multicolumn{1}{c|}{\ding{55}} & \multicolumn{1}{c|}{\ding{51}} & \multicolumn{1}{c|}{\ding{55}} & \ding{55} \\ \hline
% ROW 36 — white
\cite{EdgeAssistedVideoAnalytics} & 2023 & \multicolumn{1}{c|}{\ding{55}} & \ding{51} & \multicolumn{1}{c|}{\ding{55}} & \ding{55} & \multicolumn{1}{c|}{\ding{51}} & \multicolumn{1}{c|}{\ding{51}} & \ding{55} & \multicolumn{1}{c|}{\ding{55}} & \multicolumn{1}{c|}{\ding{55}} & \multicolumn{1}{c|}{\ding{55}} & \ding{51} & \multicolumn{1}{c|}{\ding{55}} & \multicolumn{1}{c|}{\ding{51}} & \multicolumn{1}{c|}{\ding{55}} & \ding{55} \\ \hline
% ROW 37 — grey  [NEW — Roy et al., SoCC 2024]
\rowcolor{rowgray}
\cite{hiddenCarbonServerless} & 2024 & \multicolumn{1}{c|}{\ding{51}} & \ding{55} & \multicolumn{1}{c|}{\ding{51}} & \ding{55} & \multicolumn{1}{c|}{\ding{51}} & \multicolumn{1}{c|}{\ding{51}} & \ding{55} & \multicolumn{1}{c|}{\ding{55}} & \multicolumn{1}{c|}{\ding{55}} & \multicolumn{1}{c|}{\ding{51}} & \ding{55} & \multicolumn{1}{c|}{\ding{55}} & \multicolumn{1}{c|}{\ding{55}} & \multicolumn{1}{c|}{\ding{55}} & \ding{51} \\ \hline
% ROW 38 — white  [NEW — Awwad et al., SESAME 2025]
\cite{estimatingCarbonServerless} & 2025 & \multicolumn{1}{c|}{\ding{51}} & \ding{55} & \multicolumn{1}{c|}{\ding{51}} & \ding{55} & \multicolumn{1}{c|}{\ding{51}} & \multicolumn{1}{c|}{\ding{51}} & \ding{55} & \multicolumn{1}{c|}{\ding{55}} & \multicolumn{1}{c|}{\ding{55}} & \multicolumn{1}{c|}{\ding{51}} & \ding{55} & \multicolumn{1}{c|}{\ding{55}} & \multicolumn{1}{c|}{\ding{55}} & \multicolumn{1}{c|}{\ding{55}} & \ding{51} \\ \hline
% ROW 39 — grey  [NEW — Lin & Shahrad, HotCarbon 2024]
\rowcolor{rowgray}
\cite{bridgingSustainabilityGap} & 2024 & \multicolumn{1}{c|}{\ding{51}} & \ding{55} & \multicolumn{1}{c|}{\ding{51}} & \ding{55} & \multicolumn{1}{c|}{\ding{51}} & \multicolumn{1}{c|}{\ding{51}} & \ding{55} & \multicolumn{1}{c|}{\ding{55}} & \multicolumn{1}{c|}{\ding{55}} & \multicolumn{1}{c|}{\ding{51}} & \ding{55} & \multicolumn{1}{c|}{\ding{55}} & \multicolumn{1}{c|}{\ding{55}} & \multicolumn{1}{c|}{\ding{55}} & \ding{51} \\ \hline
% ROW 40 — white  [NEW — Sharma & Fuerst, SoCC 2024]
\cite{accountableCarbonServerless} & 2024 & \multicolumn{1}{c|}{\ding{51}} & \ding{55} & \multicolumn{1}{c|}{\ding{55}} & \ding{51} & \multicolumn{1}{c|}{\ding{51}} & \multicolumn{1}{c|}{\ding{55}} & \ding{55} & \multicolumn{1}{c|}{\ding{55}} & \multicolumn{1}{c|}{\ding{55}} & \multicolumn{1}{c|}{\ding{51}} & \ding{55} & \multicolumn{1}{c|}{\ding{55}} & \multicolumn{1}{c|}{\ding{55}} & \multicolumn{1}{c|}{\ding{55}} & \ding{51} \\ \hline
\end{longtable}
\footnotesize $^\dagger$ The peer-reviewed SoCC 2023 paper \cite{ParrotFish} addresses single-function memory optimisation only. A subsequent community extension (Parrotfish-SF) added AWS Step Functions support, but this was not part of the original published contribution and is therefore not reflected in the classification.
\end{landscape}
\setlength{\tabcolsep}{6pt}
\renewcommand{\arraystretch}{1.0}

\section{Future Research Directions}
\label{sec:futureres}

With the growing needs of cloud-native applications, serverless computing has emerged as the preferred deployment and execution model due to its resource management abstraction, faster time to market, and granular billing. FaaS puts forward a serverless compute service where a developer provides the business logic as a function and a CSP takes care of the execution lifecycle. However, the burden of effectively configuring and managing the allocated resources to a function lies with the developer and is still a complex challenge. Additionally, the interplay of resource configuration parameters and various system factors aimed at optimising the runtime performance, cost, and resource utilisation requires further investigation in the serverless environments.

% Configuration aware Scheduling
\textbf{Configuration-aware Scheduling: }The configuration of function resources such as memory and CPU allocation has a profound impact on the scheduling decisions made by the underlying platform \cite{WithGreatFreedomComesGreatOpportunity}\cite{Ensure}. These decisions encompass crucial aspects like function placement on available infrastructure, the potential for co-location with other functions, and the resulting interference that may arise in multi-tenant environments. The current serverless platforms often employ relatively simple, classic scheduling algorithms that may not fully account for the unique characteristics of serverless workloads, such as their burstiness, concurrency, and very short lifecycle. Future research should delve deeper into how different resource configuration decisions influence function placement strategies. Furthermore, the impact of resource configuration on the co-location of functions needs careful examination. On one hand, placing functions with similar resource demands or communication patterns together might enhance performance through improved locality; on the other hand, co-locating resource-intensive functions could lead to adverse interference. Existing research \cite{Lambdata} discusses that resource interference and co-location may lead to resource contention like CPU, memory, and network bandwidth that can significantly downgrade performance and violate SLOs. Therefore, function configuration-aware scheduling algorithms that are also adaptable to predict and mitigate potential co-location issues should be explored in future studies.

% Dynamic Resource Allocation and associated Factors
\textbf{Dynamic Resource Allocation Factors:} The existing FaaS platforms require developers to either select or configure the compute resources for the entire scope and lifecycle of the function. The actual resource needs of a function can fluctuate significantly based on runtime factors such as the input parameters it receives and the observed workload characteristics \cite{UCCSiddharth}\cite{OFC}. This dynamism indicates that adaptive resource configuration methods, which could adjust the allocated resources of a function based on the runtime requirements, are promising for enhancing cost and performance efficiency in FaaS. In this sense, ML techniques, such as reinforcement learning and predictive modelling, could play a crucial role in developing intelligent adaptive resource configuration mechanisms that can learn from past behaviour and anticipate future resource needs based on input parameters and workload patterns. Such techniques could lead to more efficient resource utilisation and improved application performance by ensuring that functions always have the right amount of resources when they need them. Supplemental to this, the selected language runtime of the function significantly influences its resource requirements, runtime performance, and cost \cite{cordingly2020implications}. Additionally, an affinity to a specific resource \cite{CordinglyCPU}\cite{CPUTAMS} such as CPU architecture or memory of the underlying infrastructure could also affect the configuration decisions in the serverless offerings. To this end, future research should focus on comprehensively analysing the performance characteristics of serverless functions implemented in different programming languages across various serverless platforms. In addition to this, research should explore the impact of resource affinity on serverless function performance. In particular, functions that can benefit from specific CPU features or architectures could be preferentially scheduled on nodes that provide those capabilities. Hence, investigating techniques that allow developers to express resource affinity requirements and enabling serverless platforms to honour these preferences could lead to significant performance improvements.

% Decoupled or Disaggregated Resources
\textbf{Decoupled Resource Configuration: }\rev{The question of whether decoupled resource allocation should be available in production serverless platforms is increasingly being resolved. Google Cloud Run already supports independent CPU and memory allocation, and AWS Lambda Managed Instances (announced in late 2025) \cite{awsLambdaManagedInstances} allow Lambda functions to run on selected EC2 instance types with configurable memory-to-vCPU ratios, effectively enabling decoupled control within the Lambda programming model. The research challenge, therefore, is shifting from \textit{whether} decoupled resources are available to \textit{how to optimally navigate the significantly expanded and heterogeneous configuration space} that decoupling introduces.} A few of the existing works \cite{WithGreatFreedomComesGreatOpportunity}\cite{Lachesis}\cite{DecomposingGraphs} have discussed this decoupled allocation and attempted to optimise the single resource configuration for a performance improvement. As the serverless infrastructure grows in heterogeneity, comprising different types of servers, storage, and network devices, management and configuration of them emerge as a unique challenge and opportunity for dynamic resource configuration in serverless computing. As a result, the resource search space explodes with numerous possibilities, presenting a potential to explore the relationship among different resources and their impact on function performance and cost. Additionally, the support for specialised hardware such as Graphics Processing Unit (GPU) and Field-Programmable Gate Array (FPGA) remains an open question that could further have a significant impact on resource configuration decisions for specialised workloads. This opens up a possibility for heterogeneity-aware function management that can intelligently configure functions based on available hardware \cite{Fapes}. 

% Resource Configruation for Workflows with dependent functions
\textbf{Resource Configuration for Workflows:} Serverless applications are often composed of multiple interconnected functions that form workflows. The dependencies between these functions and the patterns of data flow within the workflow can significantly impact the overall performance and resource efficiency of the application. A possible future study should examine how workflow dependencies and data flow patterns between interconnected functions can be leveraged to optimise the dynamic resource configuration of the overall application. However, the studies on workflow optimisation in serverless \cite{Costless}\cite{charmseeker} and data flow aware resource management \cite{StepConf} for serverless provide a strong foundation. This could involve developing scheduling algorithms that consider the execution order and data dependencies between functions, optimising resource allocation for functions based on the volume and frequency of data they exchange, and co-locating dependent functions on the same compute nodes to minimise network latency. Additionally, these workflows could further leverage the serverless resources of different providers, where the resource configuration decisions not only impact the overall workflow performance, but also significantly influence the runtime cost. Moreover, research should identify potential cross-platform challenges in dynamic resource configuration, such as differences in resource models, scaling policies, and monitoring capabilities of different CSPs. Thus, exploring the feasibility of vendor-agnostic approaches to resource configuration and management, perhaps through the use of standardised APIs or abstraction layers, is another important direction for future work. As a result, addressing the proposed research opportunities will lead to more efficient, performant, and cost-effective serverless applications, unlocking the full potential of this promising cloud computing paradigm.

% Sustainability-Aware Configuration
\rev{\textbf{Sustainability-Aware Configuration:} As cloud providers commit to carbon-neutral operations and increasingly expose region-level carbon intensity signals, the environmental impact of serverless function configuration is emerging as a critical research direction. Function configuration choices, specifically memory allocation, CPU provisioning, instance retention policies, and concurrency settings, directly govern the energy consumed per invocation and, combined with the carbon intensity of the underlying data centre's energy mix, determine the per-invocation carbon footprint. Recent studies \cite{hiddenCarbonServerless, estimatingCarbonServerless, bridgingSustainabilityGap, accountableCarbonServerless} have begun quantifying these relationships, demonstrating that over-provisioning and idle warm container retention introduce substantial hidden emissions that are controllable through configuration. Future research should develop configuration frameworks that jointly optimise for performance, cost, and carbon efficiency treating sustainability as a first-class objective alongside the traditional KPIs surveyed in this work. This requires new multi-objective formulations, carbon-aware profiling methodologies, and production-level benchmarks that capture the real energy behaviour of functions across diverse hardware and cloud regions. Addressing this gap is essential for FaaS to align with the growing regulatory and societal expectations around sustainable computing.}

\rev{\textbf{Configuration for Serverless 2.0 Platforms:} The serverless landscape is evolving beyond the traditional FaaS model. Container-serving platforms such as Google Cloud Run and IBM Cloud Code Engine that are collectively referred to as Serverless 2.0 in the practitioner community \cite{Femux2026}\cite{mcgeeServerless20}, extend serverless principles (scale-to-zero, pay-as-you-go, event-driven) to user-defined containers with more flexible configuration, including independent CPU and memory allocation, higher concurrency limits, and longer runtimes. Recent large-scale production characterisation of Knative-based platforms reveals that user configuration behaviour in these systems, such as the prevalence of non-zero minimum scale settings and diverse concurrency limits \cite{Femux2026}, differs substantially from classical FaaS assumptions. Future research on function resource configuration must broaden its scope to encompass these richer configuration spaces, where the interaction between concurrency, minimum instances, and resource allocation introduces new optimisation challenges that existing single-function and workflow configuration methods are not designed to address.}

\section{Conclusions}
\label{sec:conclusions}

In this paper, we have presented a comprehensive literature survey on existing function configuration techniques. We propose a taxonomy that identifies various elements associated with the dynamic function configuration and its management within the broader scope of different function resources, available serverless platforms, and key performance indicators. We further discuss the three key categories of function configuration and management to analyse the existing works using the proposed taxonomy. This survey emerges as the initial study that presents a clear view on the configuration of serverless functions and aids developers to fine-tune their application performance and cost as well as CSPs to offer a true \textit{serverless} experience. Further, this work provides a concrete reference point for researchers exploring resource allocation, key performance indicators, and function configuration and management schemes in the serverless domain to advance the field. Finally, we identified existing challenges of dynamic function configuration and lay out future research directions for further research efforts. 

\bibliographystyle{ACM-Reference-Format}
\bibliography{references}

\end{document}